\documentclass[11pt]{article}
\usepackage[margin=1in]{geometry}
\usepackage[T1]{fontenc}
\usepackage{xcolor}
\usepackage{amsmath,amssymb,booktabs,graphicx,longtable}
\usepackage{newtxtext,newtxmath}
\usepackage[hidelinks]{hyperref}
\title{TopoHyperClinicalGraph: A Geometry and Topology-Aware Multimodal Framework for Cancer Survival Prediction}
\author{Noaman Khan\textsuperscript{*} \and Zubair Shah\textsuperscript{*}}
\date{}
\begin{document}
\color{black}
\maketitle
\begin{center}
College of Science and Engineering, Hamad Bin Khalifa University, Doha, Qatar\\
\textsuperscript{*}Corresponding authors\\
Noaman Khan: \href{mailto:nokh88609@hbku.edu.qa}{nokh88609@hbku.edu.qa}\\
Zubair Shah: \href{mailto:zshah@hbku.edu.qa}{zshah@hbku.edu.qa}
\end{center}

\begin{abstract}
\textbf{Background:} Accurate cancer survival prediction requires integrating tissue architecture with patient-specific clinical information. Although graph-based computational pathology models capture spatial relationships among tissue regions, conventional Euclidean representations may inadequately characterize complex tissue organization, while topological structure and clinical dependencies often remain underexplored. We propose \textbf{TopoHyperClinicalGraph}, a multimodal survival prediction framework that integrates hyperbolic tissue geometry, persistent spatial topology, and clinical graph learning to investigate their complementary contributions to prognostic modeling.

\textbf{Methods:} Our framework employs a staged learning strategy that separates geometric representation learning, topological refinement, and clinical integration. Whole-slide images are represented using pretrained UNI2-h features, reduced from 1,536 to 1,024 dimensions through training-fold principal component analysis. A spatial graph backbone combines hyperbolic neighborhood construction with Lorentz-centroid attention pooling to model tissue organization and generate patient-level survival predictions. Persistent homology summarizes multiscale spatial structure from up to 64 tissue landmarks per slide, providing complementary descriptors for a lightweight survival-logit correction. A five-node clinical graph subsequently models relationships among age, sex, pathological stage, T classification, and N classification. The staged architecture enables evaluation of each component without retraining the pathology backbone.

\textbf{Results:} We evaluated TopoHyperClinicalGraph using five-fold patient-level validation across eight cancer cohorts from TCGA and CPTAC, comprising 2,364 patients and 3,168 whole-slide images. On TCGA, the proposed framework achieved concordance indices of \textbf{0.756 in COADREAD, 0.622 in HNSC, 0.676 in BRCA, and 0.641 in STAD}, compared with 0.739, 0.600, 0.654, and 0.618, respectively, for the pathology-only spatial PatchGCN comparator. Across CPTAC cohorts, concordance indices ranged from 0.598 to 0.806. Ablation analyses indicated that geometric tissue modeling accounted for most improvements in pathology-based prediction, whereas persistent topology provided smaller, cohort-dependent effects. Clinical graph fusion outperformed its matched node-wise control in five of eight cohorts, highlighting the potential value, but also the variability, of explicitly modeling clinical relationships.

\textbf{Conclusions:} TopoHyperClinicalGraph provides a unified yet modular approach to multimodal cancer survival prediction by combining geometric tissue representations, persistent topological descriptors, and structured clinical information. The findings suggest that hyperbolic spatial modeling can strengthen pathology-based prognostic representations, while topological and clinical contributions depend on cohort characteristics. Further independent validation is required to establish generalizability and clinical utility.
\end{abstract}


\section{Background}
Whole-slide images (WSIs) provide a detailed view of tissue samples, showing cancer cells and their interactions with nearby tissues. They can also be used to study cancer prognosis by examining how tissue features relate to patient outcomes. Quantitative studies have related nuclear characteristics and tubular architecture to breast tumor grade \cite{petushi}, and histological patterns and their spatial distribution to prognostic signatures and gene expression \cite{popovici}. Associations between tissue composition, the tumor microenvironment and survival have been investigated in lung cancer \cite{racz}. Other approaches learn cellular representations without individual cell segmentation \cite{cui} or identify unsupervised histological phenotypes associated with progression-free survival in rectal cancer \cite{rectal}. This body of work motivates the study of tissue organization as a prognostic signal. Its usefulness must nevertheless be assessed within each disease and endpoint, because a pattern associated with outcome in one cohort may not retain the same meaning in another.

Survival analysis uses an observed time and an event indicator. When follow-up ends before the event, the patient contributes a right-censored observation rather than a fully observed event time. DeepSurv models nonlinear covariate effects within a Cox proportional hazards framework \cite{deepsurv}, while discrete-time neural models estimate conditional hazards over successive intervals \cite{gensheimer}. Both allow censored observations to contribute to fitting. Whole-slide studies also require patient-level grouping: several slides from one patient share one survival outcome and must remain in the same evaluation partition. This grouping determines which observations can be used to fit preprocessing transformations. A separate source of optimism arises when validation outcomes select a network checkpoint. Restricting preprocessing to training data does not remove the effects of validation-driven model selection \cite{varma}.

Patch-based processing makes gigapixel images manageable, but leaves the problem of combining thousands of local measurements into a patient-level prediction. Multiple-instance learning represents a slide or patient as a bag of patches with a shared outcome label. Parallel computing has made it easier to analyze large collections of histopathology images \cite{pmil}. Attention-based pooling identifies important tissue patches and combines their features into a slide-level representation \cite{ilse}. Clustering-based attention has also been used for whole-slide classification with limited annotations \cite{clam}, while similar methods support survival prediction \cite{yao}. However, these approaches do not directly capture spatial relationships between patches. Graph models address this limitation by connecting neighboring patches, while topological methods describe tissue organization across different spatial scales.

The choice of patch features also affects model performance. Earlier studies used convolutional networks trained on natural images \cite{xufeatures}, while newer pathology foundation models, including CTransPath, UNI, and Virchow, provide tissue-specific representations \cite{ctranspath,uni,virchow}. UNI2-h produces 1,536-dimensional patch embeddings that can be extracted and reused across experiments \cite{uni2}. This allows different graph and pooling methods to be compared using the same features. However, image preparation, artifacts, and patch sampling can still influence the results \cite{artifact}.

Graph neural networks model tissue structure by treating patches as nodes and connecting nearby patches through edges. PatchGCN uses graph convolution and attention pooling for whole-slide survival prediction \cite{patchgcn}. Other approaches model relationships between tissue regions \cite{lee} or use hypergraphs to capture higher-order interactions \cite{hgsurv}. Related survival models integrate multiple molecular data types \cite{multiomics}. Graph construction, feature propagation, and pooling each influence the final prediction. Previous studies have shown that larger neighborhoods do not always improve performance \cite{neighbors} and that simpler patch-based models can provide useful comparisons \cite{singlepatch}. These findings highlight the need to evaluate spatial context while accounting for tissue size and sampling density.

Hyperbolic geometry provides another way to represent complex relationships in data. Early work on Poincar\'e embeddings and the Lorentz model established methods for learning hierarchical representations \cite{poincare,lorentz}. These ideas were later extended to graph convolution, attention, and combined Euclidean--hyperbolic learning \cite{hgcn,hyperattention,gil}. H2-Surv also explored hyperbolic representations for multimodal cancer survival prediction \cite{h2surv}. In our approach, hyperbolic distance is used to construct spatial neighborhoods, and Lorentz-centroid attention combines patch features. A corresponding Euclidean model is used for comparison.

Persistent homology describes how connected regions and loops change across different distance thresholds \cite{edelsbrunner}. Its stability properties and numerical representations, including persistence landscapes and images, have been studied in earlier work \cite{stability,landscapes,images}. When applied to patch coordinates, persistent homology captures spatial patterns such as gaps and disconnected regions, rather than directly identifying biological structures. Topological features have also been investigated in prostate tissue diagnosis \cite{phprostate} and lung tumor subtyping from MALDI mass-spectrometry imaging \cite{maldi}. These studies motivate examining whether spatial topology provides additional information for survival prediction beyond the features learned by graph models.

WSI-GTFE combines graph representations with Mapper-based analysis of tissue architecture \cite{wsigtfe}. TopoSlide uses persistent homology to characterize the spatial arrangement of histologically defined patch clusters \cite{toposlide}. Building on this work, we focus on a staged survival model and two related questions. First, do hyperbolic neighborhood selection and Lorentz-centroid pooling improve the representation learned from fixed pathology foundation-model features? Second, do patient-level persistence descriptors add information after the geometric image model is frozen? Separating the pathology stages permits the descriptor contribution to be examined without changing the image encoder during correction fitting. A control-only correction provides the corresponding reference for assessing persistence beyond tissue size and sampling. We additionally ask whether separately recorded clinical variables improve the resulting patient prediction, and whether modeling relationships between those variables adds value over a matched node-wise clinical encoder. This question differs from adding a second representation of the same image: clinical records supply age, sex and tumor-staging measurements that are acquired separately from patch embeddings.

In this work, we developed TopoHyperClinicalGraph to integrate tissue morphology, persistent spatial topology and clinical-variable relationships for patient-level survival prediction. Its pathology branch uses hyperbolic spatial neighborhoods, real-valued graph propagation and Lorentz-centroid attention pooling. Pre-extracted UNI2-h features are reduced from 1,536 to 1,024 dimensions by training-fold principal component analysis. A 40-parameter persistence-and-control correction is fitted after the image model is frozen. The resulting pathology branch is then held fixed while a separate clinical graph learns relationships among age, sex, stage, T and N and produces four residual survival logits. The clinical graph uses Euclidean operations, and its edges are fitted without outcome labels using training-patient associations. This staged design permits comparisons with pathology-only prediction, clinical-only prediction and a parameter-matched fusion control without cross-variable message passing. We evaluate the selected configuration across TCGA-COADREAD, HNSC, BRCA and STAD and CPTAC-HNSC, PDA, CCRCC and LUAD: 2,364 patients and 3,078 slides in five patient-level folds per cohort. CPTAC-HNSC adapts a TCGA-initialized image backbone; other CPTAC backbones are trained within their cohorts. All reported scores are validation-selected. The contribution is the coordinated integration and component evaluation of these representations, rather than a new mathematical definition of hyperbolic geometry or persistence.

\section{Results}\label{sec:results}
\subsection{Primary performance across TCGA and CPTAC}
Table~\ref{tab:main_panels} organizes the primary findings into three panels. Panel (a) compares spatial PatchGCN with the pathology-only model, termed TopoHyperPatchGCN, on the four TCGA cohorts. Both use the same UNI2-h features and patient partitions; TopoHyperPatchGCN adds hyperbolic tissue neighborhoods, Lorentz-centroid pooling, and the persistence-and-control correction. Panel (b) reports this pathology-only model on the four CPTAC cohorts. Panel (c) reports the selected final multimodal model, TopoHyperClinicalGraph, which additionally uses the five-node clinical graph. The two names therefore identify distinct configurations rather than interchangeable labels for one score. Machine-readable clinical fold scores and the completed ablation inventory are provided in Additional files 2 and 3.

\begin{table}[htbp]\centering\small
\caption{Primary results organized by model and data resource. All values are five-fold mean validation-selected C-index $\pm$ sample SD. (a) TCGA pathology-only comparison. (b) CPTAC pathology-only results. (c) Final multimodal results. Bold values identify the proposed models within their panels, not statistically significant differences. All models use UNI2-h image features; panel (c) additionally uses clinical variables.}\label{tab:main_panels}\label{tab:tcga}\label{tab:cptac}
\textbf{(a) TCGA: spatial PatchGCN versus pathology-only TopoHyperPatchGCN}\\[3pt]
\begin{tabular*}{0.98\linewidth}{@{\extracolsep{\fill}}lcc}\toprule
Cohort & Spatial PatchGCN & TopoHyperPatchGCN\\\midrule
COADREAD & $0.7386\pm0.0793$ & $\mathbf{0.7544\pm0.0769}$\\
HNSC & $0.5998\pm0.0152$ & $\mathbf{0.6126\pm0.0244}$\\
BRCA & $0.6535\pm0.0604$ & $\mathbf{0.6618\pm0.0555}$\\
STAD & $0.6176\pm0.0650$ & $\mathbf{0.6213\pm0.0689}$\\
\bottomrule\end{tabular*}
\par\vspace{8pt}
\textbf{(b) CPTAC: pathology-only TopoHyperPatchGCN}\\[3pt]
\begin{tabular*}{0.98\linewidth}{@{\extracolsep{\fill}}lc}\toprule
Cohort & TopoHyperPatchGCN\\\midrule
CPTAC-HNSC & $0.5790\pm0.1214$\\
CPTAC-PDA & $0.5954\pm0.0709$\\
CPTAC-CCRCC & $0.7700\pm0.1253$\\
CPTAC-LUAD & $0.6265\pm0.0690$\\
\bottomrule\end{tabular*}
\par\vspace{8pt}
\textbf{(c) TCGA and CPTAC: final multimodal TopoHyperClinicalGraph}\\[3pt]
\begin{tabular*}{0.98\linewidth}{@{\extracolsep{\fill}}llc}\toprule
Resource & Cohort & TopoHyperClinicalGraph\\\midrule
TCGA & COADREAD & $\mathbf{0.7563\pm0.0765}$\\
 & HNSC & $\mathbf{0.6220\pm0.0226}$\\
 & BRCA & $\mathbf{0.6755\pm0.0608}$\\
 & STAD & $\mathbf{0.6409\pm0.0423}$\\
\addlinespace[3pt]
CPTAC & HNSC & $\mathbf{0.5984\pm0.1328}$\\
 & PDA & $\mathbf{0.6060\pm0.0639}$\\
 & CCRCC & $\mathbf{0.8055\pm0.1015}$\\
 & LUAD & $\mathbf{0.7192\pm0.1551}$\\
\bottomrule\end{tabular*}

\end{table}

The final multimodal model obtained TCGA C-indices of 0.7563 in COADREAD, 0.6220 in HNSC, 0.6755 in BRCA and 0.6409 in STAD, and CPTAC C-indices of 0.5984 in HNSC, 0.6060 in PDA, 0.8055 in CCRCC and 0.7192 in LUAD. CPTAC-HNSC used TCGA-initialized image adaptation with retained preprocessing; the other CPTAC image models were trained within their own cohorts. Clinical fitting used target-cohort training patients. These are adaptation or within-cohort validation results, not independent external testing. Small CPTAC event counts and fold variation limit interpretation. Clinical-only and matched node-wise alternatives, including cases that scored higher than fusion, are retained in the ablations; they are not substituted into the final-model panel.

\subsection{Ablation studies}\label{sec:all_ablations}
Component ablations compare geometric modeling, persistence descriptors and clinical graph learning. The same final configuration is used across cohorts. Additional File 1 and Additional file 3 provide fold-level results and protocol details; comparisons between different experimental families may use different source models or training schedules.

\subsubsection{Spatial controls and paired patient folds}
\begin{figure}[htbp]\centering
\includegraphics[width=\linewidth]{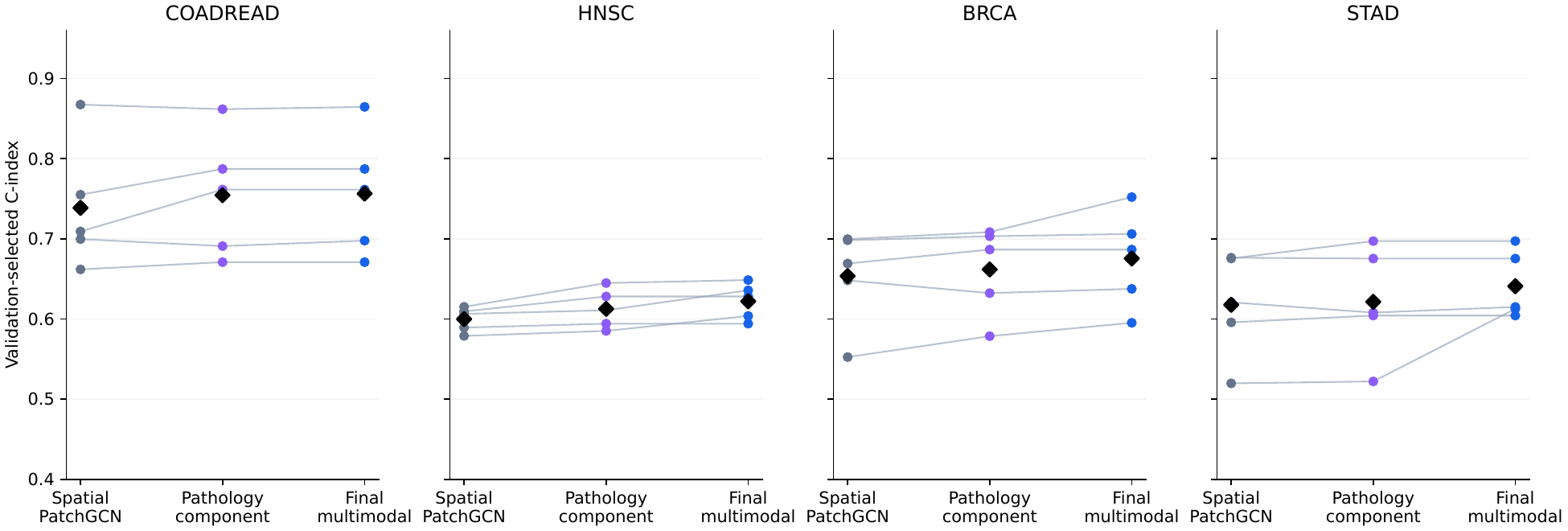}
\caption{\textbf{TCGA patient-fold comparisons.} TCGA paired-fold results for spatial PatchGCN, the frozen pathology component and final TopoHyperClinicalGraph. Lines connect the same patient fold; black diamonds show means. The final method additionally uses clinical variables. All scores are validation-selected rather than independent test estimates.}\label{fig:folds}
\end{figure}

\begin{figure}[htbp]\centering
\includegraphics[width=\linewidth]{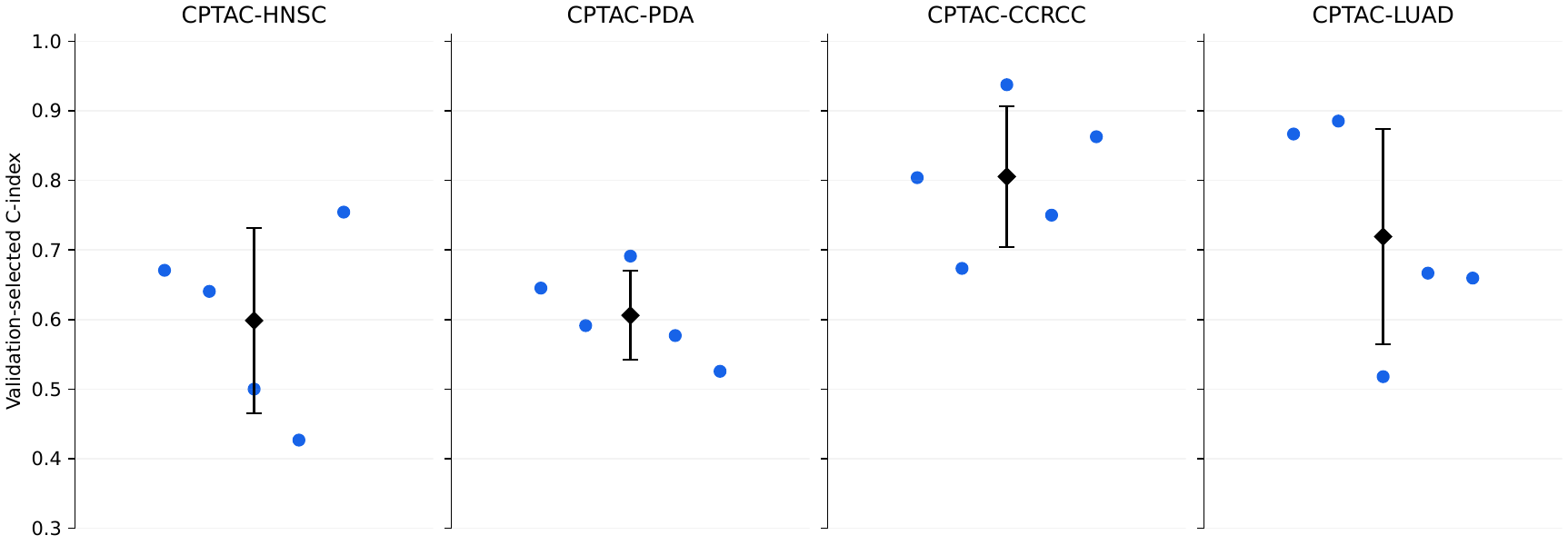}
\caption{Final TopoHyperClinicalGraph across CPTAC patient folds. Points show five folds; diamonds and error bars show mean and sample SD, not confidence intervals. Image backbones are adapted from TCGA for HNSC or trained within the other cohorts. All clinical fusion checkpoints are validation-selected; no independent test or CPTAC spatial PatchGCN comparison is implied.}\label{fig:cptac_folds}
\end{figure}

Figures~\ref{fig:folds} and \ref{fig:cptac_folds} show patient-fold results for TCGA and CPTAC, respectively. The TCGA comparison includes pathology-only spatial PatchGCN, the complete pathology model and the final multimodal model; no CPTAC spatial PatchGCN baseline was evaluated.

\subsubsection{Clinical modality and graph ablations}\label{sec:clinical_results}
TopoHyperClinicalGraph combines a clinical-variable graph with the frozen pathology component across all eight cohorts. Five clinical nodes represented age, sex, AJCC stage, T category and N category. BRCA clinical records matched 921 of the 955 patients; patients without matching records were retained using training-only imputation and missingness indicators. Public GDC clinical records matched all 1,409 patients in the other seven cohorts, although individual fields could be missing. Graph construction and clinical preprocessing used training patients only (see Methods). The pathology backbone and its persistence correction were unchanged.

In BRCA, clinical graph fusion increased mean validation-selected concordance from $0.6618\pm0.0555$ for pathology alone to $0.6755\pm0.0608$ ($\Delta=0.0137$; Table~\ref{tab:clinical_fusion}). The matched node-wise fusion control reached $0.6721\pm0.0600$, and the clinical-only graph reached $0.6675\pm0.0701$. Fusion improved four folds; the remaining fold retained the pathology prediction at epoch zero.

\begin{figure}[htbp]\centering
\includegraphics[width=\linewidth]{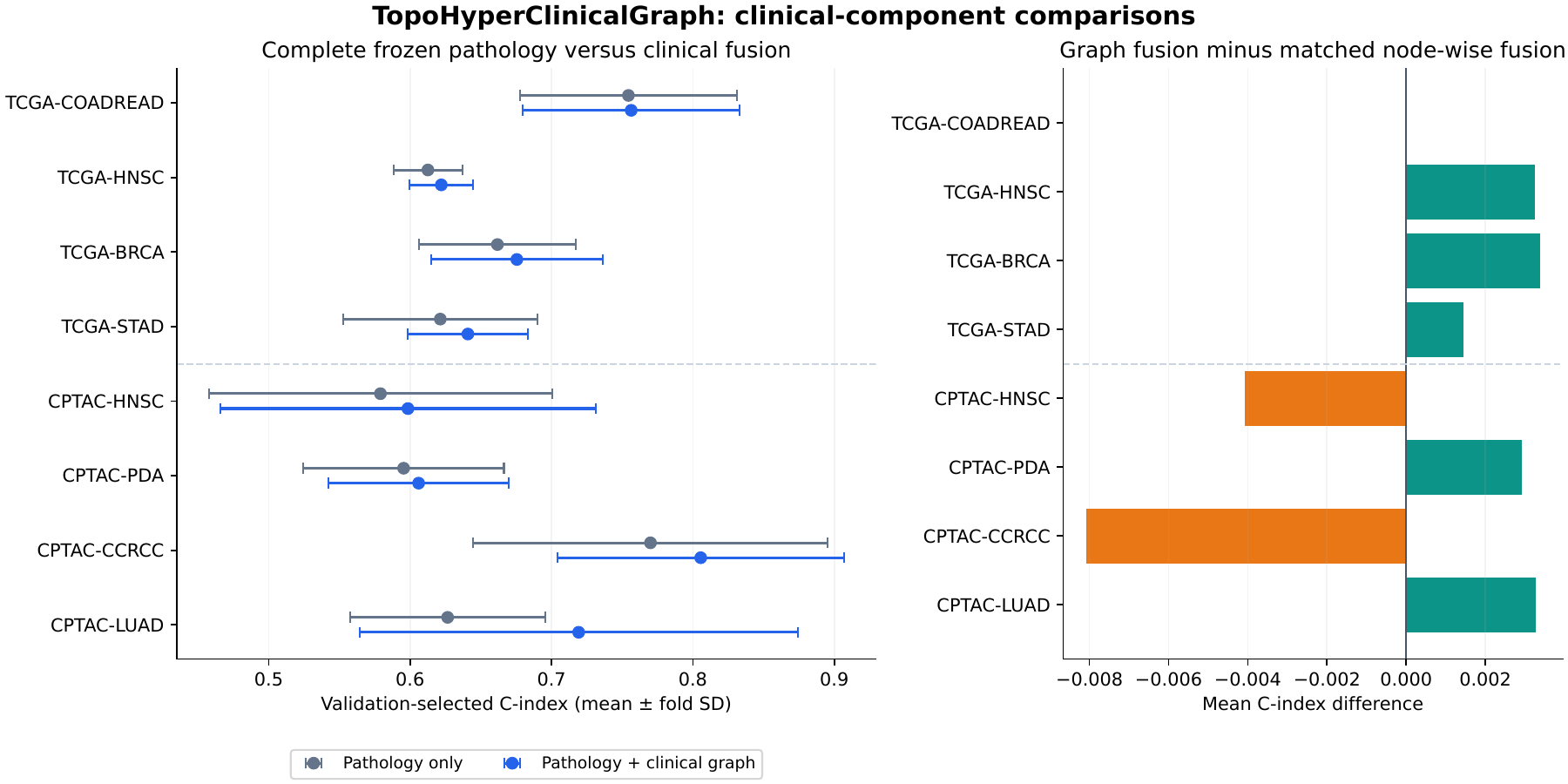}
\caption{Clinical-graph component comparisons. Left: cohort mean validation-selected C-index and fold SD for pathology alone and clinical-graph fusion. Error bars are not confidence intervals. Right: mean difference between graph fusion and the parameter-matched node-wise fusion control. Positive values favor graph message passing. No independent test evaluation or statistical superiority is implied.}\label{fig:clinical_fusion}
\end{figure}

Clinical graph fusion increased the selected mean over pathology alone in all eight cohorts, with differences of 0.0019--0.0927 (Table~\ref{tab:clinical_fusion}; Figure~\ref{fig:clinical_fusion}). It exceeded the matched node-wise control in five cohorts, tied in one and scored lower in two. Clinical-only prediction exceeded fusion in CPTAC-PDA, CPTAC-CCRCC and CPTAC-LUAD.

\begin{table}[htbp]\centering\scriptsize
\caption{Clinical-component comparisons for TopoHyperClinicalGraph. Values are five-fold mean validation-selected C-index $\pm$ sample SD. ``Pathology'' denotes the frozen pathology component, including persistence and controls. The node-wise control has identical clinical parameters but no cross-variable message passing. $\Delta$ is graph fusion minus pathology alone. Epoch-zero selection permits retention of unchanged pathology predictions.}\label{tab:clinical_fusion}
\begin{tabular}{lccccc}\toprule
Cohort & Pathology & Clinical graph & + Node-wise control & + Clinical graph & $\Delta$\\\midrule
TCGA-COADREAD & $0.7544\pm0.0769$ & $0.6746\pm0.1211$ & $0.7563\pm0.0765$ & $0.7563\pm0.0765$ & +0.0019\\
TCGA-HNSC & $0.6126\pm0.0244$ & $0.6076\pm0.0710$ & $0.6188\pm0.0207$ & $0.6220\pm0.0226$ & +0.0094\\
TCGA-BRCA & $0.6618\pm0.0555$ & $0.6675\pm0.0701$ & $0.6721\pm0.0600$ & $0.6755\pm0.0608$ & +0.0137\\
TCGA-STAD & $0.6213\pm0.0689$ & $0.6152\pm0.1160$ & $0.6394\pm0.0442$ & $0.6409\pm0.0423$ & +0.0195\\
CPTAC-HNSC & $0.5790\pm0.1214$ & $0.5809\pm0.1474$ & $0.6025\pm0.1203$ & $0.5984\pm0.1328$ & +0.0194\\
CPTAC-PDA & $0.5954\pm0.0709$ & $0.6063\pm0.0541$ & $0.6031\pm0.0651$ & $0.6060\pm0.0639$ & +0.0106\\
CPTAC-CCRCC & $0.7700\pm0.1253$ & $0.8195\pm0.0761$ & $0.8136\pm0.0992$ & $0.8055\pm0.1015$ & +0.0355\\
CPTAC-LUAD & $0.6265\pm0.0690$ & $0.7194\pm0.1285$ & $0.7159\pm0.1508$ & $0.7192\pm0.1551$ & +0.0927\\
\bottomrule
\end{tabular}

\end{table}

Because the unchanged pathology model was itself a candidate at epoch zero, selected fusion scores could not be lower than that reference on the same validation fold. This selection rule does not guarantee improvement on new patients.

\subsubsection{Pathology-component ablations: persistence and sampling controls}
Fresh inference from the frozen geometric image models gave mean C-indices of 0.7538 for COADREAD, 0.6119 for HNSC, 0.6603 for BRCA and 0.6204 for STAD. Corrections based only on tissue-size and sampling controls gave 0.7542, 0.6122, 0.6603 and 0.6214, respectively (Table~\ref{tab:topoablation}). Adding the joint persistence-and-control correction to the uncorrected image model changed the means by approximately +0.0006, +0.0007, +0.0015 and +0.0009. These increments describe the combined correction, not persistence alone.

\begin{table}[htbp]\centering\scriptsize
\caption{Descriptor ablations based on frozen HyperPatchGCN. Columns show the geometric image core, the sampling-control correction, the persistence-and-control correction (pathology component), and the extension that additionally includes cycle-curvature descriptors. Entries are five-fold mean C-index $\pm$ sample SD.}\label{tab:topoablation}
\begin{tabular}{lllll}
\toprule
Cohort & Geometry core & + Controls & + Persistence & + Curvature \\
\midrule
COADREAD & $0.7538\pm0.0775$ & $0.7542\pm0.0771$ & $0.7544\pm0.0769$ & $0.7554\pm0.0782$ \\
HNSC & $0.6119\pm0.0253$ & $0.6122\pm0.0250$ & $0.6126\pm0.0244$ & $0.6123\pm0.0244$ \\
BRCA & $0.6603\pm0.0560$ & $0.6603\pm0.0554$ & $0.6618\pm0.0555$ & $0.6615\pm0.0565$ \\
STAD & $0.6204\pm0.0691$ & $0.6214\pm0.0674$ & $0.6213\pm0.0689$ & $0.6230\pm0.0659$ \\
\bottomrule
\end{tabular}

\end{table}

To isolate persistence beyond sampling controls, we compared the pathology persistence-and-control correction and the control-only correction within the same folds (Additional File 1, Table~S1; Figure~\ref{fig:component_ablations}B). Mean differences were +0.000194 in COADREAD, +0.000363 in HNSC, +0.001480 in BRCA and $-0.000020$ in STAD. In STAD, the mean changed from 0.621355 to 0.621335. The fold-level direction varied, so these results do not establish a uniform independent persistence benefit across the evaluated cohorts.

Cycle-curvature results are included in Table~\ref{tab:topoablation} as a supplementary descriptor comparison.

\subsubsection{Hyperbolic spatial graph and pooling ablations}
The combined hyperbolic-neighborhood and Lorentz-pooling model exceeded the Euclidean control in all four TCGA cohort means (Table~\ref{tab:geometry}). The absolute differences were 0.0152 in COADREAD, 0.0121 in HNSC, 0.0061 in BRCA and 0.0028 in STAD. Propagation remained real-valued. The historical BRCA image-stage reference differs slightly from the recalculated descriptor-stage reference; source details are given in Additional File 1.

\begin{table}[htbp]\centering\small
\caption{\textbf{Geometric image-model comparison.} Historical image-stage comparison between Euclidean spatial PatchGCN and the combined hyperbolic spatial-edge/Lorentz-pooling model. Entries are five-fold mean validation-selected C-index $\pm$ sample SD. Persistence and cycle-curvature corrections are absent.}\label{tab:geometry}
\begin{tabular}{llll}
\toprule
Cohort & Euclidean & Hyperbolic & Difference \\
\midrule
COADREAD & $0.7386\pm0.0793$ & $0.7538\pm0.0775$ & +0.0152 \\
HNSC & $0.5998\pm0.0152$ & $0.6119\pm0.0253$ & +0.0121 \\
BRCA & $0.6535\pm0.0604$ & $0.6596\pm0.0563$ & +0.0061 \\
STAD & $0.6176\pm0.0650$ & $0.6204\pm0.0691$ & +0.0028 \\
\bottomrule
\end{tabular}

\end{table}

\begin{figure}[htbp]\centering
\includegraphics[width=\linewidth]{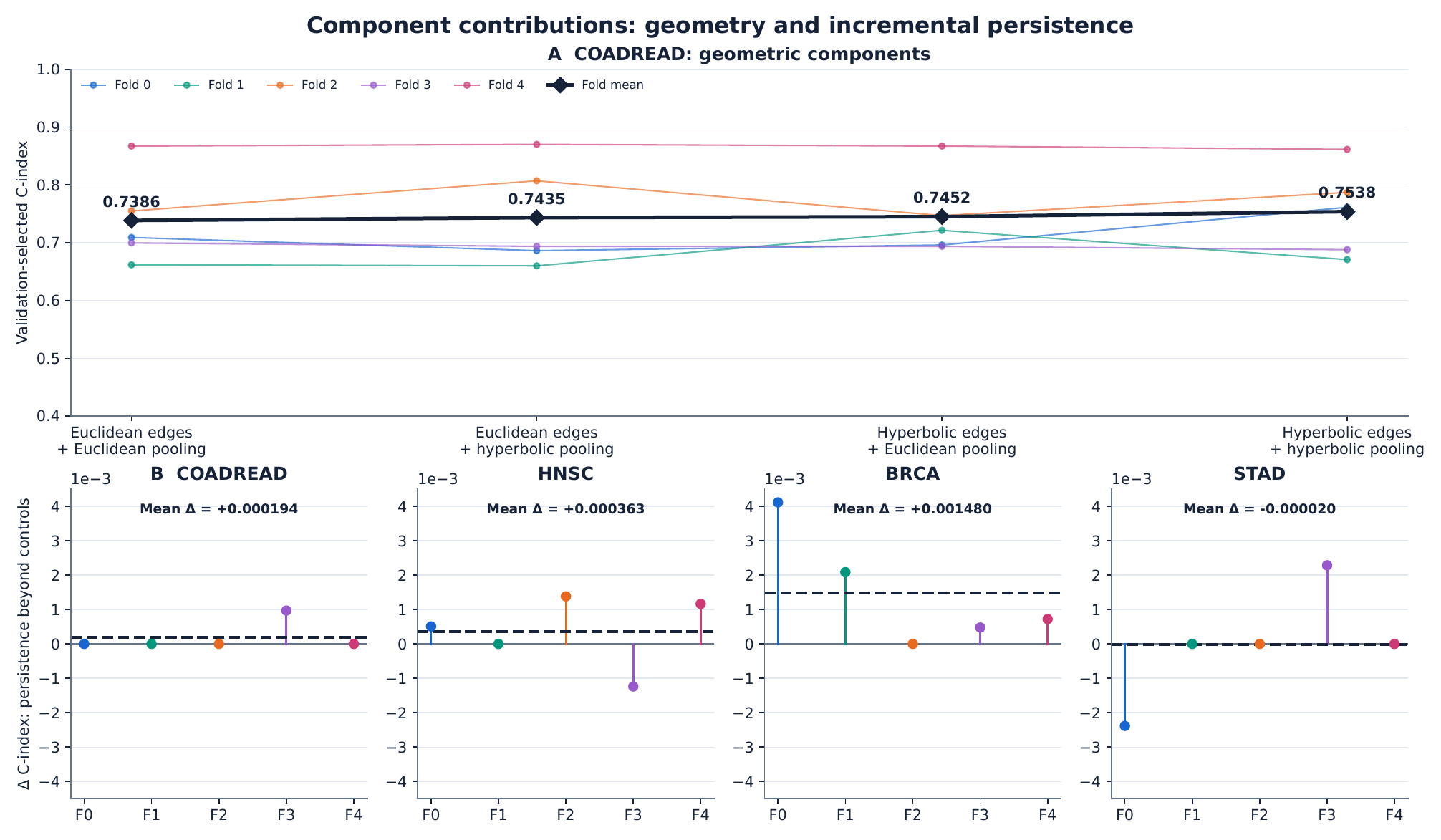}
\caption{Architectural and descriptor ablations. A: all five COADREAD folds for Euclidean edges/pooling, hyperbolic pooling alone, hyperbolic edges alone and their geometric combination; the black line shows fold means. B: paired C-index changes from adding persistence to the nuisance-only correction, using the same frozen image model within each cohort and fold. The dashed black line is the mean difference; the gray line denotes zero. These are descriptive validation-selected comparisons.}\label{fig:component_ablations}
\end{figure}

COADREAD component comparisons are shown in Figure~\ref{fig:component_ablations}A and Additional File 1, Table~S2. Hyperbolic pooling alone gave $0.7435\pm0.0906$, while hyperbolic spatial edges alone gave $0.7452\pm0.0717$. Both means exceeded the Euclidean control (0.7386); their combination gave the highest mean among these geometric configurations (0.7538). This descriptive comparison does not establish a statistically significant interaction.

Additional model comparisons and descriptor-only results are detailed in Additional File 1. The full fold-wise results are reproduced below and supplied as machine-readable tables in Additional files 2--5.

\subsection{Complete fold-wise experimental results}
The following tables retain the full experimental record in the main paper. Completed five-fold comparisons, partial runs and historical transfer evaluations are separated by protocol. Machine-readable versions remain available in Additional files 2--5.

\subsubsection{Clinical controls and multimodal models}
Table~\ref{tab:clinical_all_folds} reports all individual fold scores for pathology alone, clinical-only encoders and both fusion controls across the eight cohorts. These values correspond to Additional file 2.
\begingroup\footnotesize
\begin{longtable}{lrrrrr}
\caption{Clinical and multimodal results for every patient fold. Entries are validation-selected C-indices. Clinical MLP and clinical graph use clinical inputs alone; node fusion combines pathology with the matched node-wise control. Graph fusion is the final model.}\label{tab:clinical_all_folds}\\
\toprule Fold & Pathology & Clinical MLP & Clinical graph & Node fusion & Graph fusion\\\midrule\endfirsthead
\toprule Fold & Pathology & Clinical MLP & Clinical graph & Node fusion & Graph fusion\\\midrule\endhead
\bottomrule\endfoot
\multicolumn{6}{l}{\textbf{TCGA-COADREAD}}\\
0 & 0.7614 & 0.7157 & 0.7582 & 0.7614 & 0.7614\\
1 & 0.6709 & 0.6383 & 0.6600 & 0.6709 & 0.6709\\
2 & 0.7871 & 0.7751 & 0.8153 & 0.7871 & 0.7871\\
3 & 0.6909 & 0.5000 & 0.5000 & 0.6977 & 0.6977\\
4 & 0.8617 & 0.6657 & 0.6398 & 0.8646 & 0.8646\\
\multicolumn{6}{l}{\textbf{TCGA-HNSC}}\\
0 & 0.6109 & 0.5936 & 0.5926 & 0.6200 & 0.6357\\
1 & 0.5941 & 0.5616 & 0.5432 & 0.5941 & 0.5941\\
2 & 0.5851 & 0.6224 & 0.6369 & 0.6044 & 0.6037\\
3 & 0.6448 & 0.5767 & 0.5501 & 0.6473 & 0.6485\\
4 & 0.6281 & 0.6903 & 0.7154 & 0.6281 & 0.6281\\
\multicolumn{6}{l}{\textbf{TCGA-STAD}}\\
0 & 0.5221 & 0.6806 & 0.6961 & 0.6043 & 0.6126\\
1 & 0.6079 & 0.7935 & 0.7748 & 0.6149 & 0.6149\\
2 & 0.6042 & 0.5568 & 0.5720 & 0.6042 & 0.6042\\
3 & 0.6971 & 0.5000 & 0.5000 & 0.6983 & 0.6971\\
4 & 0.6753 & 0.5502 & 0.5333 & 0.6753 & 0.6753\\
\multicolumn{6}{l}{\textbf{CPTAC-HNSC}}\\
0 & 0.6220 & 0.5854 & 0.5488 & 0.6707 & 0.6707\\
1 & 0.6404 & 0.5169 & 0.5000 & 0.6517 & 0.6404\\
2 & 0.5000 & 0.5135 & 0.5135 & 0.5000 & 0.5000\\
3 & 0.4133 & 0.7067 & 0.5000 & 0.4533 & 0.4267\\
4 & 0.7193 & 0.8246 & 0.8421 & 0.7368 & 0.7544\\
\multicolumn{6}{l}{\textbf{CPTAC-PDA}}\\
0 & 0.6452 & 0.5806 & 0.6194 & 0.6452 & 0.6452\\
1 & 0.5766 & 0.7080 & 0.6934 & 0.5766 & 0.5912\\
2 & 0.6912 & 0.5882 & 0.5882 & 0.6912 & 0.6912\\
3 & 0.5385 & 0.5231 & 0.5538 & 0.5769 & 0.5769\\
4 & 0.5255 & 0.5474 & 0.5766 & 0.5255 & 0.5255\\
\multicolumn{6}{l}{\textbf{CPTAC-CCRCC}}\\
0 & 0.6471 & 0.8824 & 0.7843 & 0.8235 & 0.8039\\
1 & 0.6735 & 0.5306 & 0.7959 & 0.6735 & 0.6735\\
2 & 0.9375 & 0.5417 & 0.8750 & 0.9375 & 0.9375\\
3 & 0.7292 & 0.8750 & 0.9167 & 0.7708 & 0.7500\\
4 & 0.8627 & 0.7843 & 0.7255 & 0.8627 & 0.8627\\
\multicolumn{6}{l}{\textbf{CPTAC-LUAD}}\\
0 & 0.6000 & 0.7667 & 0.7500 & 0.8667 & 0.8667\\
1 & 0.6885 & 0.7869 & 0.7869 & 0.8689 & 0.8852\\
2 & 0.5179 & 0.7500 & 0.6964 & 0.5179 & 0.5179\\
3 & 0.6667 & 0.5128 & 0.5128 & 0.6667 & 0.6667\\
4 & 0.6596 & 0.8723 & 0.8511 & 0.6596 & 0.6596\\
\multicolumn{6}{l}{\textbf{TCGA-BRCA}}\\
0 & 0.7083 & 0.7178 & 0.7405 & 0.7418 & 0.7521\\
1 & 0.5785 & 0.5083 & 0.5876 & 0.5895 & 0.5952\\
2 & 0.7032 & 0.6804 & 0.7154 & 0.7067 & 0.7061\\
3 & 0.6865 & 0.5749 & 0.5978 & 0.6865 & 0.6865\\
4 & 0.6322 & 0.6845 & 0.6961 & 0.6361 & 0.6375\\
\end{longtable}
\endgroup

\subsubsection{Completed component and alternative-model comparisons}
Table~\ref{tab:all_completed_ablations} contains the 146 completed protocol rows, with five fold scores, means and sample standard deviations. Repeated reference arms are retained and are not independent replications. The machine-readable counterpart is Additional file 3.
\begingroup
\footnotesize
\begin{longtable}{p{0.30\linewidth}rrrrrr}
\caption{Individual-fold results for all inventoried completed UNI2 protocol arms. Every row contains five patient-fold C-indices and their mean and sample SD. Family/cohort headings distinguish protocols; repeated reference rows are not independent replications. Different families need not share architecture, initialization or optimization. All scores are validation-selected.}\label{tab:all_completed_ablations}\\
\toprule Variant & Fold 0 & Fold 1 & Fold 2 & Fold 3 & Fold 4 & Mean $\pm$ SD\\\midrule\endfirsthead
\toprule Variant (continued) & Fold 0 & Fold 1 & Fold 2 & Fold 3 & Fold 4 & Mean $\pm$ SD\\\midrule\endhead
\bottomrule\endfoot
\multicolumn{7}{l}{\textbf{Spatial PatchGCN controls / TCGA-COADREAD}}\\
patchgcn official uni2 pca & 0.7092 & 0.6618 & 0.7550 & 0.6996 & 0.8674 & $0.7386\pm0.0793$\\
\multicolumn{7}{l}{\textbf{Historical geometric pathology controls / TCGA-COADREAD}}\\
patchgcn hyperbolic pooling & 0.7614 & 0.6709 & 0.7871 & 0.6880 & 0.8617 & $0.7538\pm0.0775$\\
\multicolumn{7}{l}{\textbf{Pathology persistence, controls and curvature / TCGA-COADREAD}}\\
Frozen geometric image model & 0.7614 & 0.6709 & 0.7871 & 0.6880 & 0.8617 & $0.7538\pm0.0775$\\
Image + sampling controls & 0.7614 & 0.6709 & 0.7871 & 0.6899 & 0.8617 & $0.7542\pm0.0771$\\
Image + persistence and controls & 0.7614 & 0.6709 & 0.7871 & 0.6909 & 0.8617 & $0.7544\pm0.0769$\\
Image + persistence, controls and cycle curvature & 0.7647 & 0.6709 & 0.7871 & 0.6899 & 0.8646 & $0.7554\pm0.0782$\\
Topology descriptors only & 0.7059 & 0.5497 & 0.6024 & 0.6550 & 0.6974 & $0.6421\pm0.0660$\\
\multicolumn{7}{l}{\textbf{Spatial PatchGCN controls / TCGA-HNSC}}\\
hnsc uni2 patchgcn baseline & 0.6063 & 0.5892 & 0.5788 & 0.6151 & 0.6094 & $0.5998\pm0.0152$\\
\multicolumn{7}{l}{\textbf{Historical geometric pathology controls / TCGA-HNSC}}\\
hnsc hyperbolic spatial hpool & 0.6124 & 0.5935 & 0.5823 & 0.6460 & 0.6251 & $0.6119\pm0.0253$\\
\multicolumn{7}{l}{\textbf{Pathology persistence, controls and curvature / TCGA-HNSC}}\\
Frozen geometric image model & 0.6124 & 0.5935 & 0.5823 & 0.6460 & 0.6251 & $0.6119\pm0.0253$\\
Image + sampling controls & 0.6104 & 0.5941 & 0.5837 & 0.6460 & 0.6269 & $0.6122\pm0.0250$\\
Image + persistence and controls & 0.6109 & 0.5941 & 0.5851 & 0.6448 & 0.6281 & $0.6126\pm0.0244$\\
Image + persistence, controls and cycle curvature & 0.6068 & 0.5941 & 0.5864 & 0.6448 & 0.6292 & $0.6123\pm0.0244$\\
Topology descriptors only & 0.4658 & 0.5058 & 0.4986 & 0.3676 & 0.5215 & $0.4719\pm0.0617$\\
\multicolumn{7}{l}{\textbf{Spatial PatchGCN controls / TCGA-BRCA}}\\
patchgcn official uni2 pca & 0.6996 & 0.5524 & 0.6982 & 0.6691 & 0.6481 & $0.6535\pm0.0604$\\
\multicolumn{7}{l}{\textbf{Historical geometric pathology controls / TCGA-BRCA}}\\
patchgcn hyperbolic pooling & 0.7047 & 0.5754 & 0.7032 & 0.6861 & 0.6286 & $0.6596\pm0.0563$\\
\multicolumn{7}{l}{\textbf{Pathology persistence, controls and curvature / TCGA-BRCA}}\\
Frozen geometric image model & 0.7047 & 0.5754 & 0.7032 & 0.6865 & 0.6315 & $0.6603\pm0.0560$\\
Image + sampling controls & 0.7042 & 0.5764 & 0.7032 & 0.6861 & 0.6315 & $0.6603\pm0.0554$\\
Image + persistence and controls & 0.7083 & 0.5785 & 0.7032 & 0.6865 & 0.6322 & $0.6618\pm0.0555$\\
Image + persistence, controls and cycle curvature & 0.7094 & 0.5764 & 0.7032 & 0.6865 & 0.6320 & $0.6615\pm0.0565$\\
Topology descriptors only & 0.5324 & 0.3584 & 0.5680 & 0.6021 & 0.5149 & $0.5152\pm0.0939$\\
\multicolumn{7}{l}{\textbf{Spatial PatchGCN controls / TCGA-STAD}}\\
patchgcn official uni2 pca & 0.5197 & 0.6208 & 0.5958 & 0.6754 & 0.6764 & $0.6176\pm0.0650$\\
\multicolumn{7}{l}{\textbf{Historical geometric pathology controls / TCGA-STAD}}\\
patchgcn hyperbolic pooling & 0.5197 & 0.6079 & 0.6042 & 0.6949 & 0.6753 & $0.6204\pm0.0691$\\
\multicolumn{7}{l}{\textbf{Pathology persistence, controls and curvature / TCGA-STAD}}\\
Frozen geometric image model & 0.5197 & 0.6079 & 0.6042 & 0.6949 & 0.6753 & $0.6204\pm0.0691$\\
Image + sampling controls & 0.5244 & 0.6079 & 0.6042 & 0.6949 & 0.6753 & $0.6214\pm0.0674$\\
Image + persistence and controls & 0.5221 & 0.6079 & 0.6042 & 0.6971 & 0.6753 & $0.6213\pm0.0689$\\
Image + persistence, controls and cycle curvature & 0.5304 & 0.6079 & 0.6042 & 0.6971 & 0.6753 & $0.6230\pm0.0659$\\
Topology descriptors only & 0.5268 & 0.5029 & 0.4932 & 0.5097 & 0.4081 & $0.4882\pm0.0464$\\
\multicolumn{7}{l}{\textbf{COADREAD graph geometry variants / TCGA-COADREAD}}\\
patchgcn euclidean message passing control & 0.6536 & 0.6890 & 0.7390 & 0.6764 & 0.8732 & $0.7262\pm0.0879$\\
patchgcn hyperbolic message passing & 0.6569 & 0.6655 & 0.7309 & 0.6977 & 0.8703 & $0.7242\pm0.0867$\\
patchgcn hyperbolic pooling & 0.6863 & 0.6600 & 0.8072 & 0.6938 & 0.8703 & $0.7435\pm0.0906$\\
patchgcn official uni2 pca & 0.6961 & 0.7215 & 0.7470 & 0.6938 & 0.8674 & $0.7452\pm0.0717$\\
\multicolumn{7}{l}{\textbf{COADREAD message-passing/pooling combinations / TCGA-COADREAD}}\\
Matched Euclidean message passing + Lorentz pooling & 0.6569 & 0.6872 & 0.8032 & 0.7016 & 0.8617 & $0.7421\pm0.0865$\\
Hyperbolic message passing + Lorentz pooling & 0.6634 & 0.6745 & 0.7831 & 0.7093 & 0.8617 & $0.7384\pm0.0833$\\
\multicolumn{7}{l}{\textbf{Contour spectral branch, Q=64 / TCGA-HNSC}}\\
Frozen spatial backbone + contour branch & 0.6073 & 0.5880 & 0.5823 & 0.6228 & 0.6275 & $0.6056\pm0.0202$\\
\multicolumn{7}{l}{\textbf{Patch-region-slide-patient hierarchy / TCGA-HNSC}}\\
Hierarchical Lorentz pooling + persistence & 0.6048 & 0.5960 & 0.5830 & 0.6482 & 0.6182 & $0.6100\pm0.0249$\\
\multicolumn{7}{l}{\textbf{Landmark sensitivity: 16 landmarks / TCGA-HNSC}}\\
Frozen geometric image model & 0.6124 & 0.5935 & 0.5823 & 0.6460 & 0.6251 & $0.6119\pm0.0253$\\
Image + sampling controls & 0.6104 & 0.5941 & 0.5830 & 0.6460 & 0.6263 & $0.6120\pm0.0251$\\
Image + persistence and controls & 0.6114 & 0.5947 & 0.5823 & 0.6467 & 0.6251 & $0.6120\pm0.0253$\\
Image + persistence, controls and cycle curvature & 0.6083 & 0.5947 & 0.5885 & 0.6460 & 0.6251 & $0.6126\pm0.0234$\\
Topology descriptors only & 0.4977 & 0.5028 & 0.5643 & 0.5136 & 0.4889 & $0.5135\pm0.0298$\\
\multicolumn{7}{l}{\textbf{Landmark sensitivity: 32 landmarks / TCGA-HNSC}}\\
Frozen geometric image model & 0.6124 & 0.5935 & 0.5823 & 0.6460 & 0.6251 & $0.6119\pm0.0253$\\
Image + sampling controls & 0.6104 & 0.5941 & 0.5830 & 0.6460 & 0.6263 & $0.6120\pm0.0251$\\
Image + persistence and controls & 0.6119 & 0.5935 & 0.5858 & 0.6460 & 0.6263 & $0.6127\pm0.0245$\\
Image + persistence, controls and cycle curvature & 0.6098 & 0.5935 & 0.5851 & 0.6473 & 0.6263 & $0.6124\pm0.0251$\\
Topology descriptors only & 0.6083 & 0.4482 & 0.5118 & 0.4592 & 0.4820 & $0.5019\pm0.0643$\\
\multicolumn{7}{l}{\textbf{Landmark sensitivity: 64 landmarks / TCGA-HNSC}}\\
Frozen geometric image model & 0.6124 & 0.5935 & 0.5823 & 0.6460 & 0.6251 & $0.6119\pm0.0253$\\
Image + sampling controls & 0.6104 & 0.5941 & 0.5837 & 0.6460 & 0.6269 & $0.6122\pm0.0250$\\
Image + persistence and controls & 0.6109 & 0.5941 & 0.5851 & 0.6448 & 0.6281 & $0.6126\pm0.0244$\\
Image + persistence, controls and cycle curvature & 0.6068 & 0.5941 & 0.5864 & 0.6448 & 0.6292 & $0.6123\pm0.0244$\\
Topology descriptors only & 0.4658 & 0.5058 & 0.4986 & 0.3676 & 0.5215 & $0.4719\pm0.0617$\\
\multicolumn{7}{l}{\textbf{Landmark sensitivity: 128 landmarks / TCGA-HNSC}}\\
Frozen geometric image model & 0.6124 & 0.5935 & 0.5823 & 0.6460 & 0.6251 & $0.6119\pm0.0253$\\
Image + sampling controls & 0.6104 & 0.5941 & 0.5830 & 0.6460 & 0.6263 & $0.6120\pm0.0251$\\
Image + persistence and controls & 0.6210 & 0.5941 & 0.5837 & 0.6485 & 0.6286 & $0.6152\pm0.0263$\\
Image + persistence, controls and cycle curvature & 0.6230 & 0.5929 & 0.5864 & 0.6473 & 0.6292 & $0.6158\pm0.0255$\\
Topology descriptors only & 0.5008 & 0.5150 & 0.5173 & 0.5458 & 0.5582 & $0.5274\pm0.0237$\\
\multicolumn{7}{l}{\textbf{CPTAC adapted pathology and descriptor controls / CPTAC-HNSC}}\\
Adapted geometric pathology model & 0.6220 & 0.6404 & 0.4865 & 0.4133 & 0.7018 & $0.5728\pm0.1188$\\
Image + sampling controls & 0.6220 & 0.6404 & 0.4865 & 0.4133 & 0.7193 & $0.5763\pm0.1238$\\
Image + persistence and controls & 0.6220 & 0.6404 & 0.5000 & 0.4133 & 0.7193 & $0.5790\pm0.1214$\\
Image + persistence, controls and cycle curvature & 0.6707 & 0.6517 & 0.5000 & 0.4133 & 0.7018 & $0.5875\pm0.1246$\\
\multicolumn{7}{l}{\textbf{CPTAC pathology and descriptor controls / CPTAC-PDA}}\\
Frozen geometric image model & 0.6452 & 0.5766 & 0.6765 & 0.5385 & 0.5182 & $0.5910\pm0.0680$\\
Image + sampling controls & 0.6452 & 0.5766 & 0.7132 & 0.5538 & 0.5182 & $0.6014\pm0.0778$\\
Image + persistence and controls & 0.6452 & 0.5766 & 0.6912 & 0.5385 & 0.5255 & $0.5954\pm0.0709$\\
Image + persistence, controls and cycle curvature & 0.6452 & 0.5766 & 0.6838 & 0.5385 & 0.5401 & $0.5968\pm0.0651$\\
Topology descriptors only & 0.5935 & 0.5255 & 0.4632 & 0.6154 & 0.3869 & $0.5169\pm0.0941$\\
\multicolumn{7}{l}{\textbf{CPTAC pathology and descriptor controls / CPTAC-CCRCC}}\\
Frozen geometric image model & 0.6275 & 0.6735 & 0.9167 & 0.7292 & 0.8627 & $0.7619\pm0.1236$\\
Image + sampling controls & 0.6275 & 0.6735 & 0.9167 & 0.7292 & 0.8627 & $0.7619\pm0.1236$\\
Image + persistence and controls & 0.6471 & 0.6735 & 0.9375 & 0.7292 & 0.8627 & $0.7700\pm0.1253$\\
Image + persistence, controls and cycle curvature & 0.6471 & 0.6735 & 0.9375 & 0.7292 & 0.8627 & $0.7700\pm0.1253$\\
Topology descriptors only & 0.3529 & 0.3265 & 0.5833 & 0.5417 & 0.5490 & $0.4707\pm0.1209$\\
\multicolumn{7}{l}{\textbf{CPTAC pathology and descriptor controls / CPTAC-LUAD}}\\
Frozen geometric image model & 0.6000 & 0.6885 & 0.5179 & 0.6667 & 0.6596 & $0.6265\pm0.0690$\\
Image + sampling controls & 0.6167 & 0.6885 & 0.5179 & 0.6667 & 0.6596 & $0.6299\pm0.0678$\\
Image + persistence and controls & 0.6000 & 0.6885 & 0.5179 & 0.6667 & 0.6596 & $0.6265\pm0.0690$\\
Image + persistence, controls and cycle curvature & 0.6000 & 0.7869 & 0.5179 & 0.6410 & 0.6596 & $0.6411\pm0.0981$\\
Topology descriptors only & 0.3500 & 0.4918 & 0.4821 & 0.7179 & 0.6383 & $0.5360\pm0.1441$\\
\multicolumn{7}{l}{\textbf{Clinical graph and matched node-wise controls / TCGA-BRCA}}\\
Clinical graph only & 0.7405 & 0.5876 & 0.7154 & 0.5978 & 0.6961 & $0.6675\pm0.0701$\\
Clinical node-wise control & 0.7178 & 0.5083 & 0.6804 & 0.5749 & 0.6845 & $0.6332\pm0.0880$\\
Final clinical-graph fusion & 0.7521 & 0.5952 & 0.7061 & 0.6865 & 0.6375 & $0.6755\pm0.0608$\\
Pathology + node-wise clinical control & 0.7418 & 0.5895 & 0.7067 & 0.6865 & 0.6361 & $0.6721\pm0.0600$\\
Pathology-only reference & 0.7083 & 0.5785 & 0.7032 & 0.6865 & 0.6322 & $0.6618\pm0.0555$\\
\multicolumn{7}{l}{\textbf{Clinical graph and matched node-wise controls / CPTAC-CCRCC}}\\
Clinical graph only & 0.7843 & 0.7959 & 0.8750 & 0.9167 & 0.7255 & $0.8195\pm0.0761$\\
Clinical node-wise control & 0.8824 & 0.5306 & 0.5417 & 0.8750 & 0.7843 & $0.7228\pm0.1748$\\
Final clinical-graph fusion & 0.8039 & 0.6735 & 0.9375 & 0.7500 & 0.8627 & $0.8055\pm0.1015$\\
Pathology + node-wise clinical control & 0.8235 & 0.6735 & 0.9375 & 0.7708 & 0.8627 & $0.8136\pm0.0992$\\
Pathology-only reference & 0.6471 & 0.6735 & 0.9375 & 0.7292 & 0.8627 & $0.7700\pm0.1253$\\
\multicolumn{7}{l}{\textbf{Clinical graph and matched node-wise controls / CPTAC-HNSC}}\\
Clinical graph only & 0.5488 & 0.5000 & 0.5135 & 0.5000 & 0.8421 & $0.5809\pm0.1474$\\
Clinical node-wise control & 0.5854 & 0.5169 & 0.5135 & 0.7067 & 0.8246 & $0.6294\pm0.1343$\\
Final clinical-graph fusion & 0.6707 & 0.6404 & 0.5000 & 0.4267 & 0.7544 & $0.5984\pm0.1328$\\
Pathology + node-wise clinical control & 0.6707 & 0.6517 & 0.5000 & 0.4533 & 0.7368 & $0.6025\pm0.1203$\\
Pathology-only reference & 0.6220 & 0.6404 & 0.5000 & 0.4133 & 0.7193 & $0.5790\pm0.1214$\\
\multicolumn{7}{l}{\textbf{Clinical graph and matched node-wise controls / CPTAC-LUAD}}\\
Clinical graph only & 0.7500 & 0.7869 & 0.6964 & 0.5128 & 0.8511 & $0.7194\pm0.1285$\\
Clinical node-wise control & 0.7667 & 0.7869 & 0.7500 & 0.5128 & 0.8723 & $0.7377\pm0.1343$\\
Final clinical-graph fusion & 0.8667 & 0.8852 & 0.5179 & 0.6667 & 0.6596 & $0.7192\pm0.1551$\\
Pathology + node-wise clinical control & 0.8667 & 0.8689 & 0.5179 & 0.6667 & 0.6596 & $0.7159\pm0.1508$\\
Pathology-only reference & 0.6000 & 0.6885 & 0.5179 & 0.6667 & 0.6596 & $0.6265\pm0.0690$\\
\multicolumn{7}{l}{\textbf{Clinical graph and matched node-wise controls / CPTAC-PDA}}\\
Clinical graph only & 0.6194 & 0.6934 & 0.5882 & 0.5538 & 0.5766 & $0.6063\pm0.0541$\\
Clinical node-wise control & 0.5806 & 0.7080 & 0.5882 & 0.5231 & 0.5474 & $0.5895\pm0.0713$\\
Final clinical-graph fusion & 0.6452 & 0.5912 & 0.6912 & 0.5769 & 0.5255 & $0.6060\pm0.0639$\\
Pathology + node-wise clinical control & 0.6452 & 0.5766 & 0.6912 & 0.5769 & 0.5255 & $0.6031\pm0.0651$\\
Pathology-only reference & 0.6452 & 0.5766 & 0.6912 & 0.5385 & 0.5255 & $0.5954\pm0.0709$\\
\multicolumn{7}{l}{\textbf{Clinical graph and matched node-wise controls / TCGA-COADREAD}}\\
Clinical graph only & 0.7582 & 0.6600 & 0.8153 & 0.5000 & 0.6398 & $0.6746\pm0.1211$\\
Clinical node-wise control & 0.7157 & 0.6383 & 0.7751 & 0.5000 & 0.6657 & $0.6590\pm0.1030$\\
Final clinical-graph fusion & 0.7614 & 0.6709 & 0.7871 & 0.6977 & 0.8646 & $0.7563\pm0.0765$\\
Pathology + node-wise clinical control & 0.7614 & 0.6709 & 0.7871 & 0.6977 & 0.8646 & $0.7563\pm0.0765$\\
Pathology-only reference & 0.7614 & 0.6709 & 0.7871 & 0.6909 & 0.8617 & $0.7544\pm0.0769$\\
\multicolumn{7}{l}{\textbf{Clinical graph and matched node-wise controls / TCGA-HNSC}}\\
Clinical graph only & 0.5926 & 0.5432 & 0.6369 & 0.5501 & 0.7154 & $0.6076\pm0.0710$\\
Clinical node-wise control & 0.5936 & 0.5616 & 0.6224 & 0.5767 & 0.6903 & $0.6089\pm0.0508$\\
Final clinical-graph fusion & 0.6357 & 0.5941 & 0.6037 & 0.6485 & 0.6281 & $0.6220\pm0.0226$\\
Pathology + node-wise clinical control & 0.6200 & 0.5941 & 0.6044 & 0.6473 & 0.6281 & $0.6188\pm0.0207$\\
Pathology-only reference & 0.6109 & 0.5941 & 0.5851 & 0.6448 & 0.6281 & $0.6126\pm0.0244$\\
\multicolumn{7}{l}{\textbf{Clinical graph and matched node-wise controls / TCGA-STAD}}\\
Clinical graph only & 0.6961 & 0.7748 & 0.5720 & 0.5000 & 0.5333 & $0.6152\pm0.1160$\\
Clinical node-wise control & 0.6806 & 0.7935 & 0.5568 & 0.5000 & 0.5502 & $0.6162\pm0.1193$\\
Final clinical-graph fusion & 0.6126 & 0.6149 & 0.6042 & 0.6971 & 0.6753 & $0.6409\pm0.0423$\\
Pathology + node-wise clinical control & 0.6043 & 0.6149 & 0.6042 & 0.6983 & 0.6753 & $0.6394\pm0.0442$\\
Pathology-only reference & 0.5221 & 0.6079 & 0.6042 & 0.6971 & 0.6753 & $0.6213\pm0.0689$\\
\multicolumn{7}{l}{\textbf{BRCA linear clinical correction / TCGA-BRCA}}\\
Linear clinical-only model & 0.6533 & 0.6322 & 0.6255 & 0.6672 & 0.5715 & $0.6300\pm0.0366$\\
Linear clinical correction & 0.7073 & 0.5801 & 0.7032 & 0.6863 & 0.6322 & $0.6618\pm0.0546$\\
Pathology-only reference & 0.7083 & 0.5785 & 0.7032 & 0.6865 & 0.6322 & $0.6618\pm0.0555$\\
\multicolumn{7}{l}{\textbf{BRCA clinical geometry / TCGA-BRCA}}\\
Euclidean clinical-graph fusion & 0.7521 & 0.5952 & 0.7061 & 0.6865 & 0.6375 & $0.6755\pm0.0608$\\
Hyperbolic clinical aggregation fusion & 0.7495 & 0.5947 & 0.7058 & 0.6865 & 0.6375 & $0.6748\pm0.0602$\\
Hyperbolic clinical graph only & 0.7436 & 0.6122 & 0.7014 & 0.6050 & 0.6883 & $0.6701\pm0.0598$\\
Hyperbolic clinical aggregation and pooling fusion & 0.7485 & 0.5926 & 0.7055 & 0.6865 & 0.6375 & $0.6741\pm0.0605$\\
Hyperbolic clinical pooling fusion & 0.7505 & 0.5926 & 0.7055 & 0.6865 & 0.6380 & $0.6746\pm0.0611$\\
Pathology-only reference & 0.7083 & 0.5785 & 0.7032 & 0.6865 & 0.6322 & $0.6618\pm0.0555$\\
\multicolumn{7}{l}{\textbf{BRCA age hierarchy / TCGA-BRCA}}\\
Age-group feature fusion & 0.7248 & 0.5884 & 0.7049 & 0.6865 & 0.6344 & $0.6678\pm0.0557$\\
Age hierarchy, clinical-only & 0.5661 & 0.6695 & 0.6745 & 0.5000 & 0.5862 & $0.5993\pm0.0737$\\
Age-hierarchy fusion & 0.7284 & 0.5910 & 0.7046 & 0.6865 & 0.6344 & $0.6690\pm0.0556$\\
Pathology-only reference & 0.7083 & 0.5785 & 0.7032 & 0.6865 & 0.6322 & $0.6618\pm0.0555$\\
previous clinical graph & 0.7521 & 0.5952 & 0.7061 & 0.6865 & 0.6375 & $0.6755\pm0.0608$\\
\multicolumn{7}{l}{\textbf{BRCA complete clinical hierarchy / TCGA-BRCA}}\\
Complete clinical hierarchy only & 0.6587 & 0.6429 & 0.6967 & 0.6694 & 0.6243 & $0.6584\pm0.0274$\\
Permuted clinical-group control & 0.7086 & 0.5819 & 0.6693 & 0.6336 & 0.6098 & $0.6406\pm0.0497$\\
previous flat clinical graph & 0.7405 & 0.5876 & 0.7154 & 0.5978 & 0.6961 & $0.6675\pm0.0701$\\
\multicolumn{7}{l}{\textbf{BRCA clinical hierarchy fusion / TCGA-BRCA}}\\
clinical hierarchy only & 0.6587 & 0.6429 & 0.6967 & 0.6694 & 0.6243 & $0.6584\pm0.0274$\\
Fresh clinical hierarchy fusion & 0.7315 & 0.5790 & 0.7032 & 0.6865 & 0.6349 & $0.6670\pm0.0605$\\
Pathology-only reference & 0.7083 & 0.5785 & 0.7032 & 0.6865 & 0.6322 & $0.6618\pm0.0555$\\
Warm-started clinical hierarchy fusion & 0.7083 & 0.5790 & 0.7032 & 0.6865 & 0.6342 & $0.6622\pm0.0550$\\
previous flat graph fusion & 0.7521 & 0.5952 & 0.7061 & 0.6865 & 0.6375 & $0.6755\pm0.0608$\\
\multicolumn{7}{l}{\textbf{BRCA frozen report encoder / TCGA-BRCA}}\\
Pathology + clinical variables + reports & 0.7078 & 0.5832 & 0.7032 & 0.6865 & 0.6327 & $0.6627\pm0.0535$\\
Pathology-only reference & 0.7083 & 0.5785 & 0.7032 & 0.6865 & 0.6322 & $0.6618\pm0.0555$\\
Pathology + report embeddings & 0.7083 & 0.5952 & 0.7032 & 0.6865 & 0.6332 & $0.6653\pm0.0492$\\
Report-only prediction & 0.6626 & 0.6875 & 0.5558 & 0.6560 & 0.5845 & $0.6293\pm0.0562$\\
\multicolumn{7}{l}{\textbf{BRCA matched report fine-tuning / TCGA-BRCA}}\\
Text encoder last-two-layer fine-tuning & 0.7109 & 0.5785 & 0.7049 & 0.6865 & 0.6322 & $0.6626\pm0.0563$\\
Frozen text encoder, matched fusion head & 0.7114 & 0.5785 & 0.7049 & 0.6865 & 0.6322 & $0.6627\pm0.0564$\\
\multicolumn{7}{l}{\textbf{CPTAC survival/stage multitask / CPTAC-PDA}}\\
Existing pathology reference & 0.6452 & 0.5766 & 0.6912 & 0.5385 & 0.5255 & $0.5954\pm0.0709$\\
Matched survival-only head & 0.6452 & 0.6058 & 0.6912 & 0.5538 & 0.5182 & $0.6029\pm0.0692$\\
Survival + stage multitask head & 0.6452 & 0.5985 & 0.6838 & 0.5538 & 0.5182 & $0.5999\pm0.0669$\\
\end{longtable}
\normalsize

\endgroup

\subsubsection{Partial and stopped experiments}
Table~\ref{tab:partial_ablations} reports the available fold scores from incomplete or stopped protocols, corresponding to Additional file 4. Missing folds are not imputed; these rows are excluded from the primary five-fold results.
\begingroup
\footnotesize
\begin{longtable}{p{0.30\linewidth}rrrrrr}
\caption{Individual-fold scores for partial or stopped UNI2 pilots. Dashes indicate unavailable folds, not zero scores. Single-fold rows have no fold SD and are not included in primary performance comparisons. Source-fold transfer models evaluated on the same external patients are reported separately.}\label{tab:partial_ablations}\\
\toprule Variant & Fold 0 & Fold 1 & Fold 2 & Fold 3 & Fold 4 & Mean $\pm$ SD\\\midrule\endfirsthead
\toprule Variant (continued) & Fold 0 & Fold 1 & Fold 2 & Fold 3 & Fold 4 & Mean $\pm$ SD\\\midrule\endhead
\bottomrule\endfoot
\multicolumn{7}{l}{\textbf{Stopped Cayley residual pilot / TCGA-HNSC}}\\
cayley order1 & 0.6124 & -- & -- & -- & -- & $0.6124$\\
control order1 & 0.6124 & -- & -- & -- & -- & $0.6124$\\
control order2 & 0.6124 & -- & -- & -- & -- & $0.6124$\\
Frozen geometric image model & 0.6124 & -- & -- & -- & -- & $0.6124$\\
\multicolumn{7}{l}{\textbf{Single-fold spectral fusion pilot / TCGA-HNSC}}\\
fixed-fusion & 0.6327 & -- & -- & -- & -- & $0.6327$\\
hyper-only & 0.6078 & -- & -- & -- & -- & $0.6078$\\
learned-fusion & 0.6286 & -- & -- & -- & -- & $0.6286$\\
spectral-only & 0.5611 & -- & -- & -- & -- & $0.5611$\\
\end{longtable}
\normalsize

\endgroup

\subsubsection{Historical CPTAC-HNSC transfer evaluations}
Table~\ref{tab:historical_transfer} reports historical frozen source-fold predictions, corresponding to Additional file 5. Each source model was applied to the same target patients; these are not target-cohort cross-validation folds or final-method results.
\begin{table}[htbp]\centering\small
\caption{Historical frozen CPTAC-HNSC transfer pilot. Each source-fold model was applied to the same 102 target patients; these rows are not target-cohort cross-validation folds. The queue was stopped before all source models were evaluated. These values are not final-method scores.}\label{tab:historical_transfer}
\begin{tabular}{llrr}\toprule Source fold & Model & Patients & C-index\\\midrule
0 & Frozen geometric image model & 102 & 0.4536\\
0 & Frozen image + persistence correction & 102 & 0.4438\\
1 & Frozen geometric image model & 102 & 0.4905\\
1 & Frozen image + persistence correction & 102 & 0.4899\\
2 & Frozen geometric image model & 102 & 0.5211\\
2 & Frozen image + persistence correction & 102 & 0.5216\\
\bottomrule\end{tabular}
\end{table}

\section{Discussion}
TopoHyperClinicalGraph integrates learned tissue geometry, coordinate-based persistence and clinical-variable graphs. The final model achieved higher mean validation concordance than pathology-only spatial PatchGCN across the four TCGA cohorts. Pathology ablations attributed most of the architectural gain to hyperbolic neighborhoods and Lorentz-centroid pooling. The full multimodal comparison also adds clinical inputs and therefore does not isolate geometry alone.

Persistence produced small, cohort-dependent changes, including no improvement over sampling controls in STAD. Coordinate-based topology depends on tissue coverage and sampling density, and 64 landmarks may miss smaller spatial patterns. The clinical graph also showed no uniform advantage over matched node-wise fusion or clinical-only prediction. These findings support evaluating each component separately rather than assuming that additional modalities or graph operations necessarily improve prediction.

Staged training allows foundation-model features and pathology checkpoints to be reused. Only 40 parameters are fitted in the persistence correction, followed by a separately trained clinical graph. This avoids repeated image-model fitting during correction training, although overall computational efficiency was not benchmarked against other methods.

The evaluation has several limitations. Checkpoints were selected using validation outcomes, HNSC informed pathology development, and BRCA informed clinical-model selection. Epoch-zero retention ensures that selected fusion validation scores cannot fall below the pathology reference; it does not guarantee improvement on new patients. Five-fold evaluation with one initialization schedule does not provide independent uncertainty estimates. CPTAC used adaptation or cohort-specific training rather than independent testing of a locked model, and small event counts limit estimate precision.

Clinical-field availability and measurement timing require further verification. Missing records were imputed using training data, but missingness may differ in other populations. Graph-search provenance, foundation-model pretraining overlap and the historical BRCA component-reference difference also limit attribution of small effects. Calibration, demographic performance and patient-level uncertainty remain unevaluated. Independent testing and comparisons with annotated tissue structures are needed to establish predictive robustness and biological interpretation.

\section{Conclusions}
TopoHyperClinicalGraph combines hyperbolic tissue neighborhoods, Lorentz-centroid pooling, persistent spatial descriptors and clinical graph learning for whole-slide survival prediction. Five-fold experiments across four TCGA and four CPTAC cohorts showed higher selected means than the reported pathology references, with most pathology gains arising from geometric modeling and smaller, cohort-dependent contributions from persistence and clinical fusion. These validation-selected results motivate independent testing of locked models, with assessment of clinical-variable timing, calibration and uncertainty.

\section{Methods}\label{sec:methods}
\subsection{Implementation and software architecture}\label{sec:implementation}
\subsubsection{Problem formulation and architecture overview}
Patient $i$ has an observed survival outcome $(T_i,c_i)$, where $T_i$ is the time to an event or right censoring and $c_i=1$ denotes censoring ($c_i=0$ denotes an observed event). The patient contributes $m_i$ diagnostic whole-slide images and a vector of clinical measurements comprising age, sex, AJCC stage, T category and N category. Slide $j$ supplies $N_{ij}$ matched feature--coordinate pairs, stored as $X_{ij}\in\mathbb R^{N_{ij}\times1536}$ and $P_{ij}\in\mathbb R^{N_{ij}\times2}$. Coordinates are recorded in the slide's pixel coordinate system; they are not treated as calibrated physical distances. Multiple slides from one patient share one survival outcome and are not independent survival observations. The objective is to learn a patient-level risk ordering that accommodates right censoring.

\begin{figure}[htbp]\centering
\includegraphics[width=\linewidth]{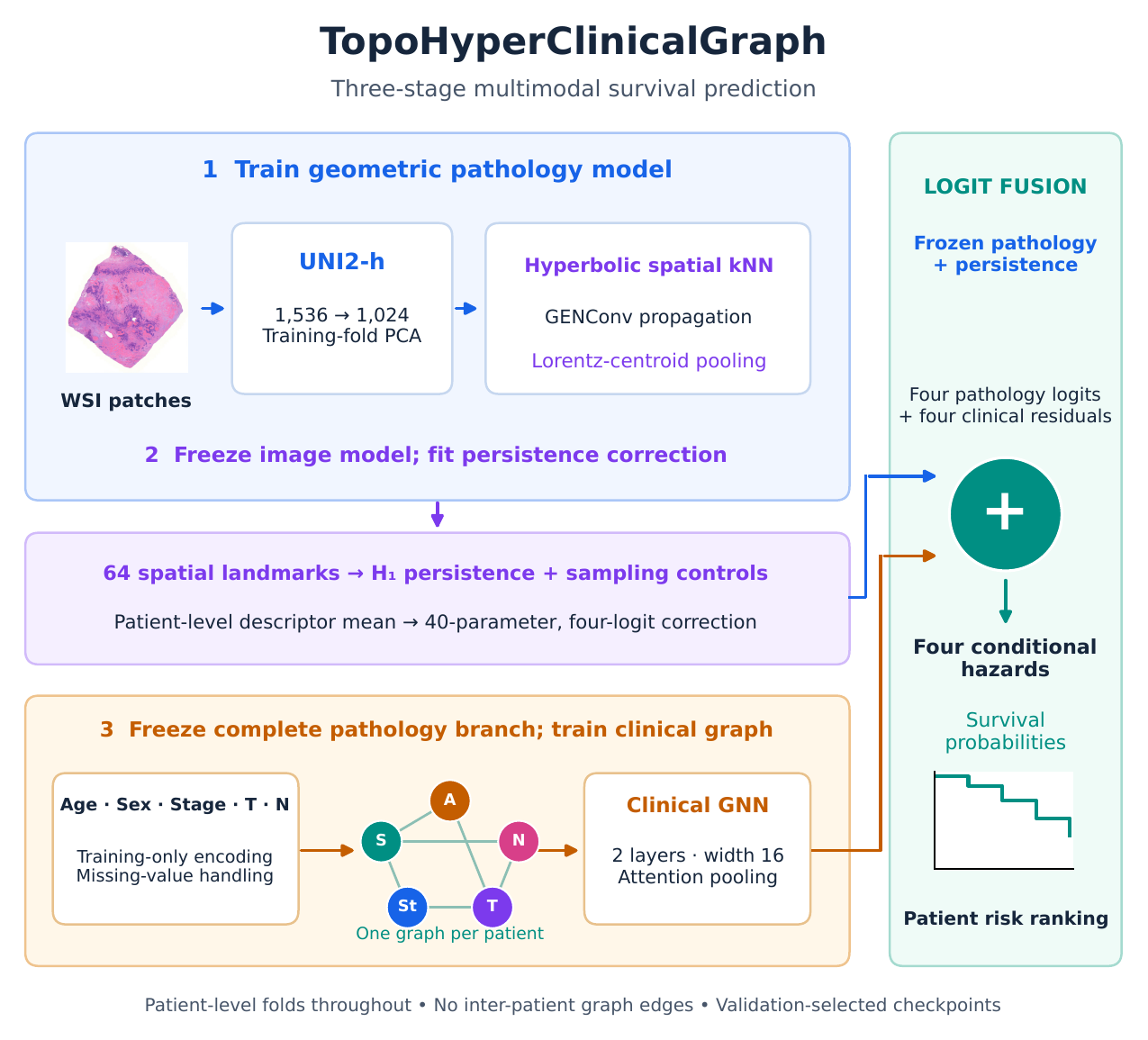}
\caption{TopoHyperClinicalGraph architecture. The pathology branch uses pre-extracted UNI2-h features, training-fold PCA, hyperbolic spatial neighborhoods, GENConv and Lorentz-centroid pooling. A frozen-image correction uses persistence and sampling descriptors. The complete pathology branch is then frozen while a separate five-node clinical graph is fitted. Clinical preprocessing and association edges use training patients only. Its four residual logits are added to the pathology logits before hazards and survival are calculated. Tissue is a real example slide; graph icons and the output curve are schematic.}\label{fig:architecture}
\end{figure}

Figure~\ref{fig:architecture} shows the three-stage TopoHyperClinicalGraph workflow. The pathology branch uses hyperbolic distances to choose spatial edges, real-valued GENConv propagation to learn contextual patch features, and Lorentz-centroid attention pooling to obtain a patient representation. Persistence summarizes coordinate organization from the same pathology slides, while clinical records provide a separate input modality.

Stage one trains the geometric image model. Its designated fold checkpoint is frozen and four pre-sigmoid logits are cached with dropout disabled. Stage two fits a 40-parameter linear correction using six persistence summaries and three sampling controls. Stage three freezes both pathology stages and fits a clinical graph that produces four additional residual logits. The sum of the pathology logits and clinical correction gives conditional hazards and a discrete survival curve. Image checkpoints follow Table~\ref{tab:training_protocols}; CPTAC-HNSC uses final-epoch adapted image weights. Persistence and clinical checkpoint selection use validation C-index. The clinical branch is trained with the residual survival objective rather than added as an independently calibrated risk score.

\begin{table}[htbp]\centering\small
\caption{Image-stage and persistence-correction protocols. Persistence corrections use 20 full-batch updates with Adam learning rate $10^{-3}$ and weight decay $10^{-3}$.}\label{tab:training_protocols}
\begin{tabular}{p{.21\linewidth}p{.22\linewidth}p{.22\linewidth}p{.23\linewidth}}\toprule
Setting & TCGA cohorts & CPTAC-HNSC & Other CPTAC cohorts\\\midrule
Image initialization & Random & Corresponding TCGA-HNSC & Random\\
PCA and time bins & Training fold & Retained TCGA fold & Training fold\\
Image epochs & 20 & 20 & 20\\
Image sampling & Unweighted & Weighted & Unweighted\\
Gradient accumulation & One patient/update & Setting 32 & One patient/update\\
Image loss $\alpha$ & 0.15 & 0 & 0.15\\
Image checkpoint & Best validation & Final epoch 20 & Best validation\\
Seed schedule & 1 at run start & 1 per fold & $1+f$ per fold\\
Correction selection & Best validation & Best validation & Best validation\\\bottomrule
\end{tabular}
\end{table}

\subsubsection{Computational reproducibility and technical stack}
Experiments used Python 3.10.20, PyTorch 2.5.1+cu124, PyTorch Geometric 2.6.1, NumPy 1.26.4, SciPy 1.13.1, scikit-learn 1.5.2 and scikit-survival 0.23.1 on Linux/WSL. Persistent homology was implemented using deterministic binary boundary reduction with integer bitsets. Image training used an NVIDIA RTX 4500 Ada GPU; clinical fitting used cached pathology logits on CPU. The complete dependency versions and environment specification are supplied with the code.

\subsubsection{Data provenance and pipeline lifecycle}
All slides from a patient remain in the same partition. Preprocessing is fitted on training patients, with retained TCGA transforms used for CPTAC-HNSC. Frozen image logits support persistence fitting, after which the complete pathology logits are cached for clinical fitting. Saved artifacts include patient partitions, preprocessing transforms, checkpoints, predictions and source hashes.

The code package includes dependency specifications, data-access instructions and verification scripts. Clean CPU checks reproduced saved concordances and correction-stage retraining across 40 folds; full image-stage retraining requires the original authorized feature and graph inputs.

AI tools were used to assist with coding and manuscript writing. Before submission, the authors will review and verify the code, results, references, and manuscript content.

\subsection{Mathematical and algorithmic framework}\label{sec:framework}
\subsubsection{Patch representations and fold-specific preprocessing}
Each tissue patch is represented by a 1,536-dimensional feature vector extracted using the UNI2-h model. The foundation model is not retrained. For each slide, feature rows and coordinate rows are paired directly; singleton batch axes are removed when present. Patch positions are not reconstructed from the number of feature rows. Spatial graph preparation uses the coordinates without access to survival labels.

The inherited image backbone expects 1024 input features. Within each TCGA training fold, randomized PCA is fitted to a sample of at most 128 patches per training slide, drawn without replacement. Validation patches use that fitted transform unchanged, and the paired model arms share it. CPTAC-PDA, CPTAC-CCRCC and CPTAC-LUAD fit PCA in the same way using their own training folds. CPTAC-HNSC instead keeps the corresponding TCGA-HNSC transform during adaptation. Sample sizes consequently vary by cohort and fold, but validation patients never contribute to PCA fitting.

The runner divides observed training-patient times at their quartiles to define four survival intervals. Censored training observations are included in these quartiles. The outer limits are set to negative and positive infinity before the intervals are applied to validation patients. CPTAC-HNSC retains its source-fold boundaries as described below. This discretization excludes validation outcomes from fitting; it does not address the separate optimism associated with selecting checkpoints on validation scores.

\subsubsection{Non-Euclidean neighborhood construction}
Each patch $v$ in slide $j$ from patient $i$ has a feature vector $x_{ijv}$ and a spatial coordinate $p_{ijv}$. Spatial edges connect each query patch to eight selected neighbors, or all other patches when fewer than eight are available. Euclidean graphs were constructed using approximate nearest-neighbor search, except for the clean BRCA graphs, which used exact search. Saved HNSC graphs were reused.

For hyperbolic neighborhood selection, coordinates are centered at the slide mean $\bar p$ and scaled by the maximum centered radius,
\begin{equation}
 R=\max\left(\max_w\|p_w-\bar p\|_2,10^{-8}\right).
\end{equation}
The centered coordinates are mapped into the Poincare ball at curvature $-1$:
\begin{equation}
 u_v=0.75\frac{p_v-\bar p}{R},\qquad
 b_v=\frac{\tanh(\|u_v\|_2)}{\|u_v\|_2}u_v.
\end{equation}
At zero norm, the mapping uses its limiting value. Distances between mapped coordinates are
\begin{equation}
 d_{\mathbb B}(b_v,b_w)=\operatorname{arcosh}\left(
 1+\frac{2\|b_v-b_w\|_2^2}{(1-\|b_v\|_2^2)(1-\|b_w\|_2^2)}
 \right).
\end{equation}
Because inverse hyperbolic cosine is monotone, the implementation ranks its argument without evaluating the function. The query patch is excluded, and the $k_v=\min(8,N_{ij}-1)$ lowest values determine its neighbors. Distances are computed in blocks to limit temporary memory use. The transformation uses only slide coordinates and does not require survival labels. Tied distances follow the backend top-$k$ ordering, which may differ across devices.

Paired Euclidean and hyperbolic graphs retain identical node features, coordinates and node ordering; only the spatial connections differ. Each recorded edge $(v,w)$ points from the query patch to its selected neighbor under PyTorch Geometric's source-to-target convention. Reverse edges and explicit self-loops are not added, so outgoing neighbor count need not equal incoming message count. For patients with several slides, node indices are offset and the graphs are combined as a disconnected union without inter-slide edges. Attention pooling then aggregates the resulting patch representations into one patient-level representation.

\begin{figure}[htbp]\centering
\includegraphics[width=\linewidth]{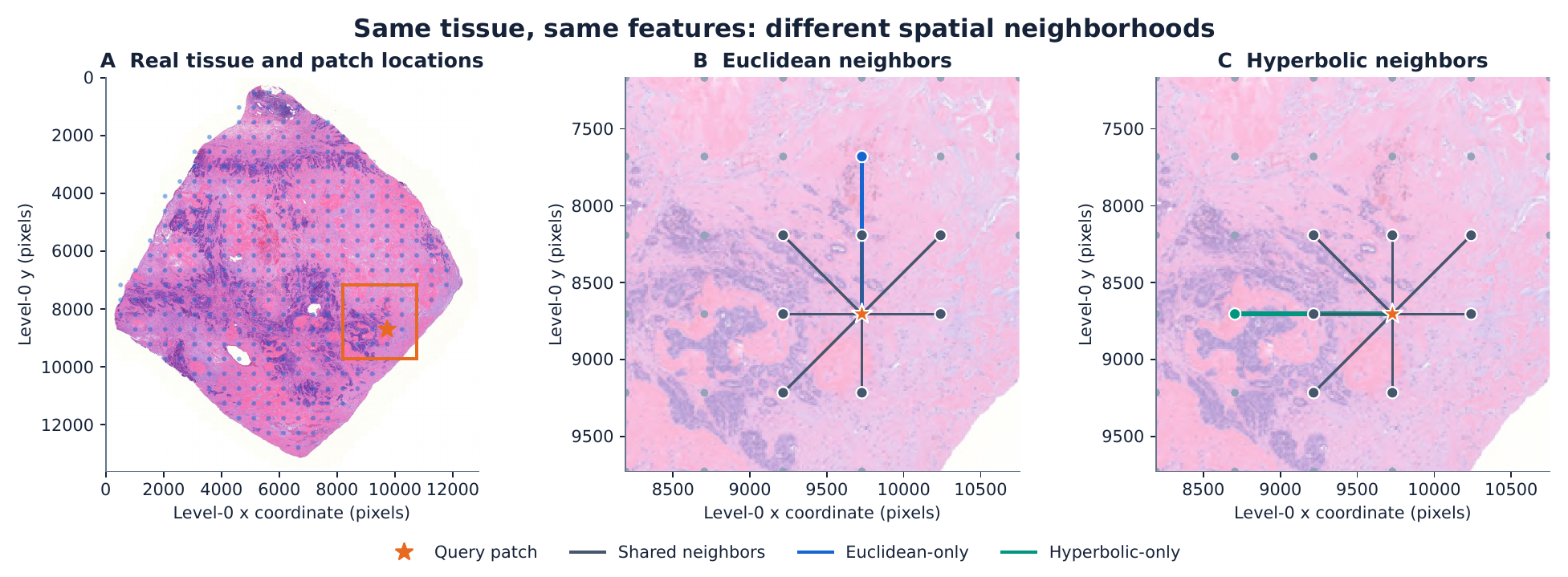}
\caption{Euclidean and hyperbolic neighborhoods on the same real TCGA-BRCA slide. Panel A shows patch locations and the crop used in B--C. Both graphs use eight selected outgoing neighbors. The query is the patch nearest the slide-coordinate centroid among patches with changed neighbor sets, selected without outcomes. Gray edges are shared; blue and green identify geometry-specific neighbors. The crop and display scale are identical. Connecting lines display neighbor membership; arrowheads are omitted for clarity.}\label{fig:spatial_graphs}
\end{figure}

Figure~\ref{fig:spatial_graphs} illustrates the two neighborhood definitions using 344 patches from one TCGA-BRCA slide. The outgoing neighbor sets differ at 66 patches, with mean neighbor-set Jaccard overlap 0.9543. This example shows the effect of the distance metric on connectivity; it does not establish a predictive benefit, and equal-distance ties can also affect selected neighbors.

\subsubsection{Message propagation and Lorentz-centroid attention pooling}
Both image models use width-128 GENConv operations on real-valued feature vectors. There is no hyperbolic message passing, geodesic parallel transport or learned manifold-valued propagation in the selected model. Its source configuration is called four-layer Patch-GCN, but it instantiates three GENConv propagation stages. Concatenating the initial projected features and the three stage outputs gives a 512-dimensional context vector for each patch. Residual graph layers and the subsequent readout modules are shared between the primary arms.

After the shared patch-feature transform, a gated network scores each vector $h_v$. With learned matrices $V,U$ and vector $w$, scores and normalized weights are
\begin{equation}
 a_v=w^\top\left[\tanh(Vh_v)\odot\sigma(Uh_v)\right],\qquad
 \alpha_v=\frac{\exp(a_v)}{\sum_w\exp(a_w)}.
\end{equation}
The Euclidean arm returns $z_E=\sum_v\alpha_vh_v$. It is an architecture-matched control using the same UNI2-h inputs. It does not reproduce the original publication's feature extractor or full training recipe.

The implemented modules are specified in Table~\ref{tab:layers}. Each GENConv uses softmax aggregation, an initial temperature of 1 with learnable temperature, and a two-layer internal network with layer normalization. The first propagation stage calls GENConv directly; the remaining two stages use residual DeepGCNLayer blocks with affine layer normalization, ReLU and dropout 0.1. No edge attributes are supplied. Gated attention uses 512-dimensional tanh and sigmoid branches, with dropout 0.25 in each branch when enabled. After pooling, a $512\rightarrow512$ linear layer with ReLU precedes the four-logit classifier. Dropout is disabled during inference and cached-logit extraction.

\begin{table}[htbp]\centering\small
\caption{\textbf{Shared image-backbone layers.} Layer specification shared by the spatial control and geometric backbone, except for the aggregation rule. Widths refer to each patch until the readout stage.}\label{tab:layers}
\begin{tabular}{lll}\toprule
Operation & Input/output width & Configuration\\\midrule
Saved UNI2-h feature & 1536 & Fixed, pre-extracted\\
Training-fold PCA & $1536\rightarrow1024$ & Randomized PCA\\
Input projection & $1024\rightarrow128$ & Linear, ReLU, dropout 0.25\\
Spatial propagation & $128\rightarrow128$ & Three GENConv stages\\
Context concatenation & $4\times128\rightarrow512$ & Projection plus three outputs\\
Patch transform & $512\rightarrow512$ & Linear, ReLU, dropout 0.25\\
Gated attention & $512\rightarrow1$ & Two 512-wide branches\\
Patient pooling & $N_i\times512\rightarrow512$ & Euclidean or Lorentz centroid\\
Readout and classifier & $512\rightarrow512\rightarrow4$ & ReLU between linear layers\\
Descriptor correction & $9\rightarrow4$ & Linear, 40 parameters\\\bottomrule
\end{tabular}
\end{table}

The geometric readout uses the same attention weights as the Euclidean control. Set $s=1/\sqrt{512}$ and $\xi_v=sh_v$, then map the tangent vector to the unit hyperboloid:
\begin{equation}
 y_v=\left(\cosh\|\xi_v\|,
 \frac{\sinh\|\xi_v\|}{\|\xi_v\|}\xi_v\right).
\end{equation}
The ambient weighted average is $\bar y=\sum_v\alpha_v y_v$. Using the inner product $\langle y,z\rangle_L=-y_0z_0+\sum_{r=1}^{512}y_rz_r$, normalize it to $\mu=\bar y/\sqrt{-\langle\bar y,\bar y\rangle_L}$. This construction gives an extrinsic Lorentz centroid rather than the exact Frechet mean. Applying the origin log map and dividing by $s$ returns a tangent vector of the original width. In the implementation, the equivalent expression uses $q=\bar y_{\mathrm{spatial}}/\bar y_0$:
\begin{equation}
 z_H=\frac{1}{s}\frac{\operatorname{atanh}\|q\|}{\|q\|}q.
\end{equation}
Series expansions handle values near zero. Nonfinite or saturated results raise errors; they are not clipped into the domain. Neither curvature nor scale is trained, and the pooling replacement adds no parameters. The propagation layers continue to operate in Euclidean feature space.

\subsubsection{Landmark selection and persistent-homology filtration}
For each slide, duplicate coordinates are removed and the remaining coordinates are sorted lexicographically. The first coordinate initializes deterministic farthest-point sampling. At each subsequent step, the algorithm selects the point with the greatest distance to its nearest selected landmark; ties use the first maximum in the sorted coordinate order. Sampling stops at $M=\min(64,n_{\mathrm{unique}})$, rather than requiring exactly 64 points for every slide. Inputs with fewer than four distinct coordinates or nonfinite coordinates are rejected.

Let $L_{ij}=\{l_1,\ldots,l_M\}\subset\mathbb R^2$ be the selected landmarks. Their distance scale is
\begin{equation}
 s_{ij}^{\mathrm{land}}=\operatorname{median}_{a}\min_{b\ne a}\|l_a-l_b\|_2.
\end{equation}
The calculation rejects a nonpositive scale. For filtration purposes, pairwise Euclidean distances are divided by $s_{ij}^{\mathrm{land}}$. These within-slide operations use neither outcomes nor data from other patients. Persistence is exact for the selected landmarks, not for the full patch-coordinate cloud.

For the normalized landmark set $Q$, the Vietoris--Rips complex at threshold $\epsilon$ is
\begin{equation}
 \mathrm{VR}_{\epsilon}(Q)=\{\sigma\subseteq Q:\max_{u,v\in\sigma}d(u,v)\leq\epsilon\}.
\end{equation}
The complete filtration is constructed through dimension two, sufficient to compute one-dimensional persistence over $\mathbb Z_2$. Edges are ordered by distance and then lexicographically; equal-scale triangles are ordered lexicographically. Boundary columns are reduced using integer-bitset exclusive-or operations. Finite pairs $(b_k,d_k)$ are retained when $l_k=d_k-b_k>10^{-12}$. An $H_1$ interval describes a loop before it is filled by triangles. The distances are normalized Euclidean coordinate distances, not hyperbolic graph shortest-path distances. Supplementary Figure~S1 illustrates the calculation on the slide used for the graph comparison.

The variable-length interval set is reduced to six summaries, not a vector of 64 sorted lifetimes: cycle count $K$, total lifetime $\sum_k l_k$, mean lifetime $\bar l$, maximum lifetime $\max_k l_k$, population lifetime standard deviation $[K^{-1}\sum_k(l_k-\bar l)^2]^{1/2}$ and persistence entropy
\begin{equation}
 E=-\sum_k q_k\log q_k,\qquad q_k=l_k/\sum_m l_m.
\end{equation}
When no positive-length intervals occur, all six summaries are set to zero. The descriptors summarize sampled spatial geometry and are not direct annotations of pathological structures. Landmark reduction bounds the problem size but may omit small-scale features; the recorded sensitivity experiment is shown in Supplementary Figure~S2.

Cycle-curvature descriptors are used only in an ablation; their construction is specified in Additional File 1.

\subsubsection{Sampling controls and linear residual correction}
The persistence summaries are accompanied by three slide-level sampling controls:
\begin{enumerate}
\item $g_{ij,1}=\log(1+N_{ij})$, the log-transformed number of patch rows;
\item $g_{ij,2}=\log(1+A_{ij})$, where $A_{ij}$ is the area of the axis-aligned coordinate bounding box;
\item $g_{ij,3}=s_{ij}^{\mathrm{land}}$, the unnormalized median nearest-neighbor distance among the selected landmarks.
\end{enumerate}
These are sampling and spatial-extent measurements. Patch count does not directly measure tissue volume, and bounding-box area can include unsampled space. The controls permit a matched comparison with persistence, but do not establish that all sampling confounding is removed.

For each slide, $t_{ij}\in\mathbb R^9$ concatenates the three controls followed by the six persistence summaries. Patient descriptors use an equal-slide mean,
\begin{equation}
 t_i=\frac{1}{m_i}\sum_{j=1}^{m_i}t_{ij}.
\end{equation}
Each component is standardized using its training-patient mean and population standard deviation; a constant component receives unit scale under StandardScaler. Validation patients do not contribute to these fitted statistics.

Stage two freezes every image-model parameter, including the survival readout, and caches the pre-sigmoid patient logits $\ell_i^{\mathrm{img}}\in\mathbb R^4$ with dropout disabled. The correction and hazards are
\begin{equation}
 \ell_i^{\mathrm{path}}=\ell_i^{\mathrm{img}}+W\widetilde t_i+b,
 \qquad \widehat h_{ik}=\sigma(\ell_{ik}^{\mathrm{path}}),
\end{equation}
where $W\in\mathbb R^{4\times9}$ and $b\in\mathbb R^4$. The total is $4\times9+4=40$ trainable parameters. Both are initialized to zero, so predictions initially equal those of the frozen image model. This low-capacity correction limits the number of new parameters; it does not guarantee prevention of overfitting.

The head receives 20 full-batch Adam updates with learning rate $10^{-3}$, weight decay $10^{-3}$, $\alpha=0.15$ and seed 1. The highest validation C-index selects its checkpoint. Unlike image training, these updates process the training patients as one batch. Each patient contributes one outcome regardless of slide count. Training logits are in-sample predictions from the image model fitted on those same training patients, not cross-fitted predictions. Validation patients contribute to neither stage's gradients. Freezing the model does not remove differences between its training and validation prediction distributions.

\subsubsection{Clinical graph construction and multimodal fusion}\label{sec:clinical_methods}
The final clinical branch evaluated in Section~\ref{sec:clinical_results} uses five variables: age, sex, AJCC stage, T category and N category. Age and the binary sex indicator (female = 1, male = 0) are imputed using training-patient medians and standardized using training-patient means and standard deviations, with separate missingness indicators. Stage, T and N use one-hot vocabularies fitted on training patients, with explicit missing and previously unseen categories. BRCA uses the locally enriched clinical table. For the other seven cohorts, public GDC case records were retrieved using an allowlist of demographic and diagnostic fields \cite{gdc_api}. Age was taken from age at diagnosis in days divided by 365.25, or age at index when a single diagnostic age was unavailable; \texttt{sex\_at\_birth} supplied the female/male indicator. Diagnoses explicitly labeled recurrent, metastatic or progressive were excluded, primary diagnoses were preferred, and conflicting stage/T/N values were left missing rather than resolved using outcomes. Clinical records matched all 1,409 patients outside BRCA. Stage was available for 282/298 COADREAD, 353/394 HNSC and 311/319 STAD patients, and 102/102 CPTAC-HNSC, 95/97 CPTAC-PDA, 94/94 CPTAC-CCRCC and 105/105 CPTAC-LUAD patients. Known-age coverage was complete except for three STAD patients. Clinical measurements and historical availability were not independently adjudicated. Survival outcomes, follow-up fields, molecular measurements and site identifiers were excluded from predictor inputs. All original patient partitions were retained, and every saved complete-pathology validation score was reproduced before fusion.

Each patient is represented by a separate five-node graph. Within a training fold, the association between two variable nodes is the maximum absolute Pearson correlation between their encoded columns; undefined correlations from constant columns are set to zero. Each variable selects its two strongest positive associations. Their union defines symmetric weighted edges, with self-loops and symmetric degree normalization. This is a statistical feature-association graph, which can reflect missingness and coding redundancy; it is not a causal graph or an inferred biological hierarchy. No edges connect patients, and validation-patient values do not determine graph construction.

Field-specific linear projections produce 16-dimensional node vectors. Two residual graph layers apply normalized adjacency aggregation, a learned linear transform, a hyperbolic tangent activation and dropout of 0.2. Attention pooling over clinical nodes feeds a zero-initialized four-output linear layer. Its output is added to the complete frozen pathology logits. This clinical graph uses real-valued operations; it does not add hyperbolic clinical message passing. Clinical-only models omit the pathology logits. A matched node-wise control replaces the clinical adjacency with the identity while retaining the same projections, parameter count, layers and pooling. Each arm uses seed 1, 100 full-batch AdamW updates, learning rate 0.001, weight decay 0.01, survival-loss $\alpha=0.15$ and gradient clipping at norm 1. Checkpoint selection uses validation C-index over epochs 0--100; epoch zero of fusion is the unchanged pathology model. Source pathology checkpoints were also validation-selected, and training-patient pathology logits are in-sample rather than cross-fitted. The final configuration was selected after exploratory BRCA comparisons and does not provide an independent test estimate.

Let $z_{iv}$ be the encoded clinical input for variable node $v$ in patient $i$, and let $\widehat A$ be the training-fold clinical adjacency after symmetrization, self-loops and symmetric degree normalization. Clinical node updates are
\begin{align}
 h_{iv}^{(0)}&=\tanh(P_vz_{iv}+b_v),\\
 h_{iv}^{(r+1)}&=h_{iv}^{(r)}+\operatorname{Dropout}\!\left[\tanh\!\left(W_r\sum_u\widehat A_{vu}h_{iu}^{(r)}+b_r\right)\right],\quad r=0,1.
\end{align}
Each hidden vector has 16 components. Clinical attention is $a_{iv}=\operatorname{softmax}_v(w_c^\top h_{iv}^{(2)}+b_c)$, and the four clinical residuals are $\delta_i=W_o\sum_v a_{iv}h_{iv}^{(2)}+b_o$. The final model is
\begin{equation}
 \boxed{\ell_i^{\mathrm{TopoHyperClinicalGraph}}=\ell_i^{\mathrm{img}}+W_p\widetilde t_i+b_p+\delta_i,\qquad
 \widehat h_{ik}=\sigma(\ell_{ik}^{\mathrm{TopoHyperClinicalGraph}}).}
 \label{eq:final_multimodal}
\end{equation}
Here $W_p\widetilde t_i+b_p$ is the already fitted persistence-and-control correction. All image and persistence parameters remain fixed while $P_v$, $W_r$, clinical attention and the output parameters are fitted. $W_o$ and $b_o$ start at zero. The clinical parameter count depends on training-fold category vocabularies; 40 parameters describes only the persistence correction, not this branch.

\subsubsection{Fixed choices and reproducible training sequence}
Experiments retain eight spatial neighbors, coordinate scale 0.75, curvature $-1$, width 128, pooling scale $1/\sqrt{512}$, 64 landmarks and four survival intervals. Input width 1024 preserves backbone compatibility. The pooling scale accounts for representation width, while coordinate scaling bounds the radius before mapping. These values define the implemented configuration; the experiments do not establish their optimality. Supplementary Figure~S2 reports the saved landmark-count comparison and preparation cost. HNSC informed development, so the configuration cannot be described as prespecified independently of all inspected outcomes.

\begin{quote}\small
\textbf{Algorithm 1: one patient-level fold of TopoHyperClinicalGraph.}
\begin{enumerate}
\item Read patient identifiers, survival labels, paired slide features/coordinates and the supplied training/validation partition; keep every patient's slides together.
\item Fit PCA and four time intervals on training patients only, or load the retained source-fold transforms for CPTAC-HNSC.
\item Map each slide's centered coordinates to the Poincare ball and select directed hyperbolic neighbors; assemble each patient's slides as a disconnected graph union.
\item Train the geometric image model for 20 epochs using the cohort-specific protocol in Table~\ref{tab:training_protocols}; load its designated checkpoint.
\item Freeze the image model, disable dropout and cache its four logits for every patient.
\item Independently select up to 64 spatial landmarks per slide; compute normalized Vietoris--Rips persistence and the six persistence and three sampling descriptors.
\item Average slide descriptors within each patient; fit descriptor means and scales on training patients only.
\item Initialize the $9\rightarrow4$ correction to zero; optimize on training patients for 20 full-batch updates and select its best validation checkpoint.
\item Freeze the persistence correction and cache the complete pathology logits; retain these unchanged predictions as the clinical stage's epoch-zero candidate.
\item Match age, sex, stage, T and N to patients; fit imputation, scaling and category vocabularies using training patients only.
\item Estimate training-only clinical associations, form the separate five-node graph for each patient, and initialize its four-output residual layer to zero.
\item Train the clinical graph for 100 full-batch AdamW updates on training patients; select the highest validation C-index over epochs 0--100.
\item Add clinical residual logits to the complete pathology logits, compute hazards, survival and patient risks, and save predictions and provenance.
\end{enumerate}
\end{quote}

\subsection{Datasets, optimization and evaluation metrics}\label{sec:evaluation}
\subsubsection{Discrete-time survival objective}
The model predicts a discrete-time survival distribution rather than ordinary four-class labels. The three quartile cuts in the relevant training-time distribution define four intervals indexed by $k\in\{0,1,2,3\}$. CPTAC-HNSC retains its TCGA source intervals. For patient $i$, logits $\ell_{ik}$ give conditional hazards $\widehat h_{ik}=\sigma(\ell_{ik})$. Survival through interval $k$ and the patient-level risk score are
\begin{equation}
 \widehat S_{ik}=\prod_{r=0}^{k}(1-\widehat h_{ir}),\qquad R_i=-\sum_{k=0}^{3}\widehat S_{ik}.
\end{equation}
Let $Y_i$ be the interval containing the observed event or censoring time, with $c_i=1$ for censoring and $c_i=0$ for an event. Event and censoring contributions are
\begin{align}
 L_{u,i}&=-(1-c_i)\{\log\widehat S_{i,Y_i-1}+\log\widehat h_{i,Y_i}\},\\
 L_{c,i}&=-c_i\log\widehat S_{i,Y_i},\\
 \mathcal L_{\mathcal B}&=\frac{1}{|\mathcal B|}\sum_{i\in\mathcal B}
 \left[L_{u,i}+(1-\alpha)L_{c,i}\right].
\end{align}
Here $\widehat S_{i,-1}=1$ for the first interval and $\mathcal B$ is the training batch. This is a weighted discrete-time negative log-likelihood objective, not Cox partial likelihood. In particular, a censored patient contributes survival through the observed interval under the implemented convention. The usual TCGA and correction settings use $\alpha=0.15$; CPTAC-HNSC image adaptation uses $\alpha=0$. Logarithm arguments are bounded for numerical stability. One patient supplies one loss contribution and one risk prediction.

\subsubsection{Cross-validation and checkpoint selection}
The existing five patient-level splits were retained. Both arms used 20 epochs, batch size one patient, Adam learning rate $2\times10^{-4}$ and weight decay $10^{-5}$. The current TCGA runner uses unweighted sampling, one optimizer update per patient and $\alpha=0.15$ for both training and validation. These settings differ from the upstream defaults of weighted sampling, accumulation over 32 patients and $\alpha=0$. Random initialization follows the existing runner, with seed 1 initialized at run start rather than a blanket claim that every fold is independently reset.

The best validation C-index within each fold determines the selected TCGA checkpoint. Image-model selection, persistence-head selection and final clinical-graph selection use best validation concordance, with the CPTAC-HNSC image exception described below. Such scores describe internal validation with model selection, not untouched test performance. They must not be interpreted as leakage-free generalization estimates simply because PCA and time bins were training-only. No claim is made that the original publication's performance has been reproduced.

Validation patients contribute neither to preprocessing fits nor to gradient updates. Checkpoints are selected on those patients, so the folds are validation partitions rather than independent test sets.

Image, persistence and clinical stages are trained sequentially. TCGA and cohort-trained CPTAC image checkpoints are validation-selected. CPTAC-HNSC instead uses its final epoch-20 adapted image weights; subsequent corrections are validation-selected. All reported fold scores therefore represent internal validation with model selection.

\subsubsection{Study design and data sources}
This retrospective computational study used existing TCGA diagnostic-slide feature files and survival metadata for COADREAD, HNSC, BRCA and STAD, together with separate CPTAC-HNSC, CPTAC-PDA, CPTAC-CCRCC and CPTAC-LUAD experiments. No new tissue collection, treatment assignment or patient recruitment was performed. TCGA metadata and slide identifiers originated from the existing Patch-GCN project and associated cohort files. CPTAC overall-survival labels and patient partitions were obtained from the Patho-Bench distribution, and paired patch features and coordinates from the official UNI2-h feature repository \cite{pathobench,uni2}. Selected slide identifiers were matched to the survival tables before feature and graph preparation. Source repositories are \url{https://github.com/mahmoodlab/Patch-GCN}, \url{https://huggingface.co/datasets/MahmoodLab/Patho-Bench} and \url{https://huggingface.co/datasets/MahmoodLab/UNI2-h-features}. Repository revisions, selected slide manifests and result-file hashes are retained with the computational artifacts; the exact source cohorts and exclusions are given in Table~\ref{tab:cohort}.

\begin{table}[htbp]\centering
\caption{Analyzed survival cohorts. Counts refer to selected patients and slide records, not all available feature files.}
\begin{tabular}{lrr}\toprule
Cohort & Patients & Slides\\\midrule
TCGA-COADREAD & 298 & 301\\
TCGA-HNSC & 394 & 414\\
TCGA-BRCA & 955 & 1019\\
TCGA-STAD & 319 & 343\\
CPTAC-HNSC & 102 & 243\\
CPTAC-PDA & 97 & 227\\
CPTAC-CCRCC & 94 & 218\\
CPTAC-LUAD & 105 & 313\\\bottomrule
\end{tabular}\label{tab:cohort}
\end{table}

One BRCA feature file was truncated and its required slide had no usable graph. Excluding that slide removed its otherwise unrepresented patient, leaving 955 of the intended 956 patients. Both model arms used the same exclusion, without reference to prediction performance. The HNSC graph diagnostic covers 472 slides, whereas survival modeling uses 414; these are different analysis sets. All three training stages completed five folds in each of the four CPTAC cohorts. BLCA has no completed predictive experiment in this study.

A survival record contains the patient identifier, observed time and censoring indicator, with $c_i=1$ for right censoring and $c_i=0$ for an event. Slides from one patient share that record and partition. For CPTAC, event/censoring counts were 33/69 in HNSC, 71/26 in PDA, 20/74 in CCRCC and 23/82 in LUAD. Selected patients had complete endpoints and consistent labels across their slides. The endpoint is overall survival, although follow-up and ascertainment may differ between data resources.

\subsubsection{Cohort survival characteristics}
The selected TCGA cohorts contained 58 observed events among 298 COADREAD patients, 176 among 394 HNSC patients, 130 among 955 BRCA patients and 120 among 319 STAD patients. The corresponding censored counts were 240, 218, 825 and 199. Median observed times were 22.32, 20.82, 28.84 and 16.53 months. These medians summarize recorded event/censoring times; they are not Kaplan--Meier median overall survival or estimates of median follow-up. Differences in event burden affect the number of comparable patient pairs and the stability of fold-level concordance. We therefore report each cohort separately rather than treating the cohorts as exchangeable validation replicates.

\subsubsection{Separate CPTAC full-method evaluation}
The CPTAC analysis uses each cohort's supplied five overall-survival patient partitions. It is reported separately from the matched TCGA comparison. Within a patient, all slide graphs form a disconnected union and remain in one partition. Every final CPTAC score uses hyperbolic spatial edges, Lorentz-centroid pooling, six persistence descriptors, three sampling controls and the same five-node clinical graph. The pathology-only component is reported separately. Curvature and multitask stage prediction are excluded. The CPTAC comparisons do not include a spatial PatchGCN baseline; clinical-only and node-wise fusion controls are evaluated within the final-model component analysis.

For CPTAC-PDA, CPTAC-CCRCC and CPTAC-LUAD, each image model was initialized independently within its own cohort, using saved UNI2-h features rather than initializing from an unrelated cancer checkpoint. Hyperbolic spatial graphs used eight neighbors, fixed curvature magnitude 1 and coordinate scale 0.75. A training-only PCA projected 1536-dimensional features to 1024 dimensions, and the four survival intervals were estimated from training patients only. The image stage used 20 epochs, Adam learning rate $2\times10^{-4}$, weight decay $10^{-5}$, unweighted patient sampling, one optimizer update per patient and survival-loss parameter $\alpha=0.15$. The seed was reset to $1+f$ for fold $f$. The best image checkpoint by validation concordance was then frozen. After independent per-slide topology computation, the 40-parameter persistence correction was fitted on training patients only, with training-only descriptor standardization, 20 full-batch epochs, Adam learning rate $10^{-3}$, weight decay $10^{-3}$, $\alpha=0.15$ and seed 1. Its best validation checkpoint provided the frozen pathology-component logits. The final clinical graph was then fitted for 100 updates using the protocol above.

CPTAC-HNSC followed a different image-initialization protocol and is identified accordingly. Each fold used its corresponding TCGA-HNSC image checkpoint, PCA and survival boundaries. The image model was adapted on that CPTAC training partition for 20 epochs using the earlier official-source training loop: weighted patient sampling, $\alpha=0$, gradient accumulation setting 32, Adam and seed 1. The source revision was \texttt{823addaee5b8f4cc2bec3ea8e5e0077b2a5115a4}. Only the final epoch-20 adapted image weights were saved. These weights were frozen before fitting the same 20-epoch persistence correction used above. The complete pathology branch was then frozen before fitting the final clinical graph for 100 updates. Thus, the HNSC result uses a final-epoch adapted image checkpoint and validation-selected persistence and clinical checkpoints; it must not be described as using best-validation image weights. Prior inspection of CPTAC-HNSC during transfer and adaptation experiments is disclosed.

Recorded days were converted to months using $365.25/12$ days per month. Patient-level concordance was used throughout. Although the supplied partition column is named ``test'', it was used to select checkpoints and therefore served as validation. Held-out fold sizes were 19--20 patients with 14--15 events for PDA, 18--19 with four events for CCRCC, 21 with four or five events for LUAD, and 20--21 with six or seven events for HNSC. These patients were excluded from preprocessing fits and gradient updates. Their use for checkpoint selection still permits optimism. The experiments assess within-cohort fitting or adaptation, not prediction in a new cohort without fitting. Supplementary Figure~S3 gives the saved trajectories for the three cohort-trained models.

Table~\ref{tab:training_protocols} summarizes the protocol differences. CPTAC-HNSC adaptation uses Adam learning rate $2\times10^{-4}$ and weight decay $10^{-5}$. Its retained TCGA time boundaries are an explicit exception to fitting new boundaries within the CPTAC training fold.

\subsubsection{Component definitions and controlled comparisons}
The baseline uses Euclidean spatial edges and Euclidean gated-attention pooling. Removing the persistence correction from the pathology component leaves the geometric core: hyperbolic edges and Lorentz-centroid pooling. A three-control correction has 16 parameters; adding the six persistence descriptors gives the selected 40-parameter correction. The curvature arm uses 16 descriptors and 68 parameters. The topology-only arm fits those same 16 descriptors without image logits. All pathology descriptor-correction arms share the patient partitions, training-only standardization, objective, learning rate, weight decay, number of updates and selection rule. The clinical stage is a separate comparison using the 100-update AdamW protocol described above. Zero initialization makes every image-augmented arm start from the same frozen image predictions.

Secondary COADREAD experiments vary spatial edges, pooling and message passing separately. Their saved results support component comparisons within that cohort. The final model does not switch components according to each cancer's highest score: it consistently uses the persistence-and-control pathology component followed by flat clinical-graph fusion. Clinical-only and node-wise alternatives are controls; curvature is reported separately.

To isolate persistence beyond sampling controls, the principal descriptor contrast is $C_{\mathrm{image+controls+persistence}}\allowbreak-C_{\mathrm{image+controls}}$ within each fold. The contrast against the uncorrected image model instead measures the joint change from persistence and controls. These two estimands are reported separately; neither is a statistical test of causality.

\subsubsection{Evaluation metrics and statistical summaries}
Patient-level discrimination is measured with scikit-survival's \texttt{concordance\_index\_censored}. For risk $R_i$, observed time $T_i$ and event indicator $e_i=1-c_i$, the ordered comparable pairs are
\begin{equation}
 \mathcal C=\{(i,j):e_i=1,\ T_i<T_j\}\ \cup\ \{(i,j):e_i=1,\ e_j=0,\ T_i=T_j\}.
\end{equation}
Using the risk-tie tolerance $\tau=10^{-8}$, concordance is
\begin{equation}
 \widehat C=\frac{\sum_{(i,j)\in\mathcal C}\left[\mathbf1\{R_i>R_j+\tau\}+\tfrac12\mathbf1\{|R_i-R_j|\leq\tau\}\right]}{|\mathcal C|}.
 \label{eq:cindex}
\end{equation}
An earlier event should receive the higher risk. The library excludes tied event--event times and includes an event/censoring pair at the same recorded time according to the set above. Without comparable pairs the score is undefined, not zero; the selected runs require defined validation concordance. Saturated hazards can produce tied risks whose ordering depends on numerical precision. The implementation and dtype are consequently part of the evaluation specification.

For $F=5$ folds, the reported mean is $\bar C=F^{-1}\sum_f C_f$ and the sample standard deviation is $s_C=[\sum_f(C_f-\bar C)^2/(F-1)]^{1/2}$. Paired differences are formed within a fold before averaging. No hypothesis tests, independent repeated-seed experiments or patient-bootstrap confidence intervals have been completed. Because training sets overlap, the five folds are not independent experimental replicates. Their SD describes variation and cannot establish superiority. Cohort means are kept separate rather than combined into a pan-cancer efficacy estimate.

\subsubsection{Implementation verification}
Tests checked simple persistence examples, coordinate invariances, agreement with independently calculated boundary ranks, frozen checkpoint loading and training-only standardization. Additional File 1 gives these checks and the historical BRCA reference re-evaluation; descriptor comparisons retain their recalculated frozen reference.

\section*{List of abbreviations}
BRCA, breast invasive carcinoma; CCRCC, clear cell renal cell carcinoma; COADREAD, colon and rectal adenocarcinoma; CPTAC, Clinical Proteomic Tumor Analysis Consortium; GCN, graph convolutional network; GDC, Genomic Data Commons; HNSC, head and neck squamous cell carcinoma; LUAD, lung adenocarcinoma; PCA, principal component analysis; PDA, pancreatic ductal adenocarcinoma; SD, standard deviation; STAD, stomach adenocarcinoma; TCGA, The Cancer Genome Atlas; WSI, whole-slide image.

\section*{Additional files}
\noindent\textbf{Additional file 1 (Additional\_File\_1.pdf): Supplementary component comparisons.} Pathology ablations, landmark sensitivity, learning histories, complete and partial experimental inventories, and supplementary figures.\par
\noindent\textbf{Additional file 2 (Additional\_File\_2.csv): Clinical model fold scores.} Machine-readable final, pathology-only and clinical-control validation scores for the eight cohorts.\par
\noindent\textbf{Additional file 3 (Additional\_File\_3.csv): Completed ablation inventory.} The 146 completed protocol rows with individual fold scores, means and sample standard deviations, including repeated reference arms.\par
\noindent\textbf{Additional file 4 (Additional\_File\_4.csv): Partial experiment inventory.} Incomplete or stopped experimental arms with their actual available fold coverage; missing folds are not imputed.\par
\noindent\textbf{Additional file 5 (Additional\_File\_5.csv): Historical transfer results.} Historical frozen CPTAC-HNSC source-fold evaluations, distinguished from target-cohort cross-validation.\par

\section*{Declarations}
\subsection*{Ethics approval and consent to participate}
This study analyzed existing, publicly available, de-identified TCGA and CPTAC datasets. No participants were recruited, and no directly identifiable patient information was accessed.
\subsection*{Consent for publication}
Not applicable. The manuscript contains no identifiable patient information.
\subsection*{Availability of data and materials}
Source data are available through the NCI Genomic Data Commons, TCIA and the referenced Hugging Face datasets, subject to their respective access and licensing terms. Derived result tables and source-file hashes accompany this draft. The source code and reproducibility materials will be made publicly available upon publication at \url{https://github.com/Noaman67khan/TopoHyperClinicalGraph}. A source archive and verification instructions are prepared for confidential editorial and reviewer assessment. The repository is not currently claimed to provide public access. Raw foundation-model features remain subject to their source licensing terms.
\subsection*{Competing interests}
The authors declare that they have no competing interests.
\subsection*{Funding}
This research received no specific funding. Noaman Khan is a recipient of a scholarship from Hamad Bin Khalifa University. Open access publication funding is provided by Qatar National Library.
\subsection*{Authors' contributions}
NK conceived the study, developed and implemented the method, conducted the experiments, and wrote the initial manuscript. ZS supervised the project, contributed to scientific discussions and methodological refinement, and reviewed and revised the manuscript. Both authors contributed to the final manuscript revisions, read and approved the final manuscript, and agreed to its submission.
\subsection*{Acknowledgements}
The authors acknowledge the patients and contributing institutions of TCGA and CPTAC, and the developers of the source models and datasets.

\end{document}